
\input harvmac

\font\teneufm=eufm10
\font\seveneufm=eufm7
\font\fiveeufm=eufm5
\newfam\eufmfam
\textfont\eufmfam=\teneufm
\scriptfont\eufmfam=\seveneufm
\scriptscriptfont\eufmfam=\fiveeufm
\def\frak#1{{\fam\eufmfam\relax#1}}

\font\teneusm=eusm10
\font\seveneusm=eusm7
\font\fiveeusm=eusm5
\newfam\eusmfam
\textfont\eusmfam=\teneusm
\scriptfont\eusmfam=\seveneusm
\scriptscriptfont\eusmfam=\fiveeusm

\font\tenmsx=msam10
\font\sevenmsx=msam7
\font\fivemsx=msam5
\font\tenmsy=msbm10
\font\sevenmsy=msbm7
\font\fivemsy=msbm5
\newfam\msafam
\newfam\msbfam
\textfont\msafam=\tenmsx  \scriptfont\msafam=\sevenmsx
  \scriptscriptfont\msafam=\fivemsx
\textfont\msbfam=\tenmsy  \scriptfont\msbfam=\sevenmsy
  \scriptscriptfont\msbfam=\fivemsy

\def\msbm#1{{\fam\msbfam\relax#1}}
\def\title#1#2#3#4{\nopagenumbers\abstractfont\hsize=\hstitle
\rightline{ESENAT-#1}
\rightline{YUMS-#2}%
\vskip 1in\centerline{\titlefont #3}\abstractfont\vskip .2in
\centerline{{\titlefont#4}}\vskip 0.3in\pageno=0}
\def\jsp{\centerline{Jae-Suk Park
\footnote{$^\dagger$}{e-mail: pjesenat@krysucc1.bitnet}
}\bigskip\centerline{{\it ESENAT Theoretical Physics Group}}
\centerline{{\it Seodaemun P.O.~Box 126, Seoul 120-600, Korea}}
\centerline{{\it and}}
\centerline{{\it Institute for Mathematical Science}}
\centerline{{\it Yonsei University, Seoul 120-749, Korea}}
\bigskip}
\def\abs#1#2{\centerline{{\bf Abstract}}\vskip 0.2in
{#2}\Date{#1}}
\def\acknow#1{\bigbreak\bigskip\noindent{\it Acknowledgements.} {#1}}
\def\fonote#1{\foot{#1}}

\def\a{\alpha}    \def\b{\beta}       \def\c{\chi}       \def\d{\delta}
    \def\e{\varepsilon} \def\f{\phi}       \def\F{\Phi}
               \def\l{\lambda}
\def\L{\Lambda}   \def\m{\mu}                 \def\r{\rho}
  \def\o{\omega}      \def\O{\Omega}     \def\p{\psi}
\def\P{\Psi}            \def\S{\Sigma}     
        \def\w{\varphi}    

%

\def\CA{{\cal A}} \def\CB{{\cal B}}  
\def\CD{{\cal D}} \def\CE{{\cal E}}  \def\CF{{\cal F}}  \def\CG{{\cal G}}
   \def\CL{{\cal L}}

\def\CW{{\cal W}} \def\CM{{\cal M}}
\def\CX{{\cal X}} \def\CY{{\cal Y}}
\def\CJ{{\cal J}} \def\CK{{\cal K}}
%
\def\rd{\partial}

\def\darr#1{\raise1.5ex\hbox{$\leftrightarrow$}\mkern-16.5mu #1}
\def\Ha{{1\over2}}
\def\ha{{\textstyle{1\over2}}}
\def\fr#1#2{{\textstyle{#1\over#2}}}
\def\Fr#1#2{{#1\over#2}}
\def\tr{\hbox{Tr}\,}

\def\roughly#1{\raise.3ex\hbox{$#1$\kern-.75em\lower1ex\hbox{$\sim$}}}
\def\ato#1{{\buildrel #1\over\longrightarrow}}
%
\def\cmp#1#2#3{Commun.\ Math.\ Phys.\ {{\bf #1}}, {#3} {(#2)}}
\def\pl#1#2#3{Phys.\ Lett.\ {{\bf #1}}, {#3} {(#2)}}
\def\np#1#2#3{Nucl.\ Phys.\ {{\bf #1}}, {#3} {(#2)}}
\def\pr#1#2#3{Phys.\ Rev.\ {{\bf #1}}, {#3} {(#2)}}

\def\ijmp#1#2#3{Int.\ J.\ Mod.\ Phys.\ {{\bf #1}}, {#3} {(#2)}}

\def\jdg#1#2#3{J.\ Differ.\ Geom.\ {{\bf #1}}, {#3} {(#2)}}
\def\pnas#1#2#3{Proc.\ Nat.\ Acad.\ Sci.\ USA.\ {{\bf #1}}, {#3} {(#2)}}
\def\top#1#2#3{Topology {{\bf #1}}, {#3} {(#2)}}
\def\zp#1#2#3{Z.\ Phys.\ {{\bf #1}}, {#3} {(#2)}}
\def\prp#1#2#3{Phys.\ Rep.\ {{\bf #1}}, {#3} {(#2)}}

\def\ptrsls#1#2#3{Philos.\ Trans.\  Roy.\ Soc.\ London
{{\bf #1}}, {#3} {(#2)}}

\def\am#1#2#3{Ann.\ Math.\ {{\bf #1}}, {#3} {(#2)}}

\def\im#1#2#3{Invent.\ Math.\ {{\bf #1}}, {#3} {(#2)}}
\def\plms#1#2#3{Proc.\ London Math.\ Soc.\ {{\bf #1}}, {#3} {(#2)}}
\def\dmj#1#2#3{Duke Math.\  J.\ {{\bf #1}}, {#3} {(#2)}}

\def\jgp#1#2#3{J.\ Geom.\ Phys.\ {{\bf #1}}, {#3} {(#2)}}

\def\adE{\hbox{\rm Ad}(E)}
\def\pr{\prime}
\def\ppr{{\prime\prime}}
\def\Dp{\rd_A}
\def\Dpp{\bar\rd_A}
\def\bs{{\bf s}}
\def\bbs{{\bf\bar s}}
\def\hW{\widehat W}
\def\dw{\delta_{\!{}_{W}}}
\def\dh{\delta_{\!{}_{H}}}
\def\BC{\msbm{C}}
\def\BP{\msbm{P}}
\def\BE{\msbm{E}}
\def\BR{\msbm{R}}
\title{93-01}{93-10}{$N=2$ Topological Yang-Mills
Theory}{on Compact K\"{a}hler Surfaces}
\bigskip
\jsp
\bigskip
\bigskip
\abs{April; November, 1993}{We study a topological Yang-Mills
theory with $N=2$ fermionic symmetry.
Our formalism is a field theoretical interpretation of the
Donaldson polynomial invariants on compact K\"{a}hler surfaces.
We also study an analogous theory on compact oriented Riemann surfaces
and briefly discuss a possible
application of the Witten's non-Abelian localization formula to
the problems in the case of compact K\"{a}hler surfaces.
\bigskip
\medskip
\centerline{ To appear in Comm.~Math.~Phys.}}

%

\newsec{Introduction}
Several years ago, Witten introduced
the topological Yang-Mills theory (TYMT) [1]
on general $4$-manifolds to provide a quantum field theoretical
interpretation of the Donaldson polynomial invariants [2].
The basic property of the TYMT is that there is a fermionic
symmetry which localizes the path integral to an integral over the
moduli space $\CM$ of anti-self-dual (ASD) connections.
Geometrically, the fermionic operator $\dw$ acts on $\CM$ as
the exterior derivative. The action functional
of the TYMT can be written as an $\dw$-exact form,
\eqn\biia{
S_W = \dw V\;.
}
In the TYMT, correlation functions of physical observables
correspond to the Donaldson polynomial invariants.

The moduli space $\CM$ of ASD connections on a compact K\"{a}hler surface
has natural complex and K\"{a}hler structures [3], which implies that
the TYMT has actually $N=2$ fermionic symmetry generated by
the holomorphic and the anti-holomorphic parts of $\dw$,
i.e. $\dw = \bs +\bbs$.
Geometrically,
we can interpret $\bbs$ as the Dolbeault
cohomology operator on $\CM$.
Then, the $N=2$ version of the topological
action may be written as
\eqn\biib{
S= \bs\bbs{\bf B_{T}}\;.
}
In the first part of this paper, we study TYMT on compact
K\"{a}hler surfaces with $N=2$ fermionic symmetry.
In Sect.2, we briefly sketch Donaldson theory
and the TYMT of Witten in order to
make this paper reasonably self-contained and to set up notations
for the later sections.
In Sect.3, we construct the $N=2$ TYMT on compact K\"{a}her surfaces.
After a short discussion on the anti-self-duality relative to
K\"{a}hler form, we show that the theory has $N=2$ fermionic symmetry
thereby
deriving the corresponding algebra. We construct
topological actions and study zero-modes of fermionic fields
which naturally represent the cohomology structures of $\CM$.
In Sect.4, we study topological observables and their correlation
functions which can be interpreted as the Donaldson invariants on
compact K\"{a}hler surfaces.
As an example, we show that one of the correlation functions
can be identified with the symplectic volume of the moduli space
under some favorable conditions.
This may be compared with
the differential geometrical approach of Donaldson (sketched in p.~294--295
of [2]). Note that the main application of the Donaldson's
original work [2] was to the differential topology of complex
algebraic surfaces. This is because, over compact K\"{a}hler surfaces,
one can use algebro-geometric techniques which enhance the
computability of the invariants.
Although it is not clear that our treatment will lead to some explicit
expressions for the invariants, we may hope to illuminate our
understanding of Donaldson theory in the field theoretical
interpretation.

Recently, Witten obtained a general expression for
the two-dimensional analogue of the Donaldson invariants [4, 5].
In the second part of this paper, we discuss some analogies
between the $N=2$ TYMT on compact K\"{a}hler surfaces and the TYMT
on compact Riemann surfaces (Sect.3 of [5])
In Sect.5.1, we study an $N=2$ version
of the TYMT on compact oriented Riemann surface as a straightforward
application of the techniques developed in the first part of
this paper. In Sect.5.2, we give a basic description of the
Donaldson's algebro-geometrical approach to his invariants [2].
In Sect.5.3, we suggest that the non-Abelian localization formula
of Witten [5] can be applied to Donaldson theory on compact K\"{a}hler
surfaces.

We understand that Galperin and Ogievetsky studied a
TYMT on compact K\"{a}hler surfaces, as well as the hyperK\"ahler
case, with extended fermionic symmetry [6].
Their construction is based on the twisting
of the $N=2$ super-Yang-Mills theory [1].

\newsec{Preliminary}
In this section, we will briefly sketch the Donaldson invariants
[2] and the TYMT of Witten [1].
A comprehensive exposition of Donaldson theory can be found in [7].
It will be also useful to consult with [8].
A detailed introduction of topological field theories in general
can be found in [9].

\subsec{Sketch of the Donaldson Invariants}
Let $M$ be a smooth oriented compact Riemann four-manifold and
let $E$ be a vector bundle over $M$ with a reduction of the structure
group to $SU(2)$. Let $\adE$ be the Lie algebra bundle associated to $E$
by the adjoint representation. The bundle $E$ is classified by
the instanton number $k=\left<c_2(E), [M]\right>$.
Throughout this paper,
we will assume that $k$ is strictly positive.
Let $\CA$ be the space of all connections
on $E$ and $\CG$ be the group of gauge transformations.
Since any two connections differ by an element of $\adE$-valued
one form,  $\CA$ is an affine space and its tangent space $T\CA$
consists of $\adE$-valued one form on $M$.
We define a quotient space $\CB=\CA/\CG$, the set of gauge equivalence
classes of connections. We also introduce $\CB^* = \CA^*/\CG$,
where $\CA^*$ denotes the space of irreducible connections.
Due to Atiyah and Singer [10], there exists an universal
bundle $\BE$ over the product space $M\times\CB^*$ as an adjoint
bundle.  Let $A^{m,n}(M\times\CB^*)$ denote the space of
$\hbox{\rm Ad}(\BE)$-valued $m$-forms on $M$ and $n$-forms on $\CB^*$
respectively.
Let $\CF$ be the
total curvature two-form, which can be decomposed into
components
\eqn\kaa{\eqalign{
\CF^{2,0} &= dA + A\wedge A\;,\cr
\CF^{1,1} &= \dh A\;,\cr
\CF^{0,2} &= -G_A[\dh A,*\dh A]\;,\cr
}
}
where $G_{\!A} = (d_{\!A}*d_{\!A})^{-1}$ and $\dh A$ represents
tangent vectors on $\CB^*$, i.e. $\dh A$ is the horizontal projection
of $\adE$-valued one-form $\d A$ ($d_A^*\dh A =0$).

One can  interpret the operator $\dh$ as the exterior
covariant derivative (the horizontal part of exterior
derivative $\d$ on $\CA$) on $\CB^*$ [11],
\eqn\kab{\eqalign{
&d_A: A^{m,n}(M\times\CB^*)\rightarrow A^{m+1,n}(M\times\CB^*)\;,\cr
&\dh: A^{m,n}(M\times\CB^*)\rightarrow A^{m,n+1}(M\times\CB^*)\;.\cr
}}
Then, one can find that
\eqn\kac{
\dh A = \CF^{1,1}\;,\qquad \dh \CF^{1,1} = -d_A\CF^{0,2}\;,\qquad \dh
\CF^{0,2} =0\;.
}
Note that $\dh$ is nilpotent up to a gauge transformation
generated by $\CF^{0,2}$, i.e. $\dh^2 A = -d_A\CF^{0,2}$.
Provided that $\dh$
acts on gauge invariant functionals, $\dh$ can be interpreted as
the de Rham cohomology operator on $\CB^*$. That is,
$\dh$ is an operator of the  equivariant cohomology.
The de Rham cohomology classes on $\CB^*$ can be easily
constructed from the characteristic class
$c_2(\BE) = \Fr{1}{8\pi^2}\tr \CF^2$
which is a closed form of degree $2$ on $M\times\CB^*$,
\eqn\yya{
(d + \dh)\Fr{1}{8\pi^2}\tr\CF^2 =0\;.
}
We expand $c_2(\BE)$ as follows;
\eqn\yyb{
\Fr{1}{8\pi^2}\tr\CF^2 =\sum_{r=0}^{4}\CW_{r}{}^{4-r}  \qquad
\hbox{where}\quad\CW_{r}{}^{4-r}\in A^{r,4-r}(M\times\CB^*)\;,
}
and define
$W_r{}^{4-r}\equiv \int_M \CW_{r}{}^{4-r}\wedge O_{4-r}$
for a harmonic $4-r$ form $O_{4-r}$ on $M$.
Equation \yya\
shows that $W_r{}^{4-r}$ is an element of de Rham cohomology
classes on $\CB^*$,
\eqn\yyc{
\dh W_r{}^{4-r} =0\;,\qquad W_r{}^{4-r}\in H^{4-r}(\CB^*)\;,
}
which depends only on the cohomology class of $O_{4-r}\in H^{4-r}(M)$.
In the Poincar\'e dual picture, one can define
$W_r{}^{4-r}\equiv \int_{Y_r} \CW_{r}{}^{4-r}$ in terms of
a $r$-dimensional homology
cycle $Y_r$ which  is Poincar\'e dual to $O_{4-r}$.

Let $\CM$ be the moduli space of ASD connections -- the set of gauge
equivalence classes of ASD connections.
For a given ASD  connection  $A$, a neighborhood of the point
$[A]$   in  the  moduli  space $\CM$ of ASD connections should
satisfy the following equations;
\eqn\easd{\eqalign{
&F^+(A+\d A)=d^+_A\d A + (\d A\wedge\d A)^+ =0\;,\cr
&d_A^*\d A = 0\;,\cr}
}
where $F^+$  denotes the self-dual part of the curvature and $*$
is the Hodge star operator.  The second equation in \easd\
restricts the infinitesimal variation $\d A$ to the direction
orthogonal to the pure gauge variation.
The linearisation of Eq.\easd\ gives rise to
an elliptic operator $\d_A = d^+_A\oplus d^*_A$ which
leads to the instanton complex of Atiyah-Hitchin-Singer [12];
\eqn\eic{
0\;\ato{}\; A^0(\adE)\;\ato{d_A}\; A^1(\adE)\;\ato{d^+_A }\;
A^{2}_{+}(\adE)\;\ato{{}}\; 0\;.
}
The cohomology groups of this elliptic complex are
\eqn\ecoh{\eqalign{
&H^0_A=\hbox{Ker }\triangle^0\;,\qquad\triangle^0=d_A^*d_{A}\;,\cr
&H^1_A=\hbox{Ker }\triangle^1\;,\qquad
\triangle^1=d_A d_{A}^* +d_A^{+*}d_A^+\;,\cr
&H^2_A=\hbox{Ker }\triangle^2\;,\qquad\triangle^2=d_A^+d_A^{+*}\;,\cr}
}
where $h^i_A=\hbox{dim } H^i_A$ denotes the $i$-{\it th} Betti number.
The zeroth cohomology is trivial when $A$ is irreducible
and the first cohomology group can be identified with the
tangent space at $[A]$ on $\CM$.  The singularities in $\CM$
arise if $h^{0}_{A} \neq 0$ or $h^{2}_{A} \neq 0$.
The formal
dimension of $\CM$, for a simply connected $M$, is given by
the index of the instanton complex,
\eqn\eind{
\hbox{ind}(\d_A) =  h^1_A -h^0_A -h^2_A = 8k - 3(1 + b^+(M))\;,
}
where $b^+(M)$ is the dimension
of the self-dual harmonic two-forms on $M$.

Let $M$ be a simply connected, oriented $4$-manifold. Then, the
essential cohomological data are contained in $H^2(M)$. Thus, we
have a de Rham cohomology on $\CB^*$,
\eqn\sasha{
W_2{}^{2}=\Fr{1}{4\pi^2}\int_M \tr\left(\CF^{0,2} \CF^{2,0}
+\ha\CF^{1,1}\wedge\CF^{1,1}\right)
\wedge
O_2\;,
}
which depends only on the cohomology class of $O_2\in H^2(M)$
or on the homology class of $\S \in H_2(M)$ which is
Poincar\'e dual to
$O_2$. Equivalently, $W_{2}{}^{2}$ defines the Donaldson $\m$-map,
$\m(\S)$.
The basic idea of Donaldson is to use the moduli space $\CM^*$
of irreducible ASD connections as a fundamental homology cycle
in order to evaluate a cup product of the cohomology classes
$\m(\S_i)$ for $\S_i\in H_2(M)$,
\eqn\cupp{
\left<\m(\S_1)\cup\cdots\cup\m(\S_d), [\CM^*]\right>\;.
}
In oder to have a differential topological interpretation of
the above parings, further specifications are required.
That is, the moduli space $\CM^*$ should be a smooth oriented
$2d$-dimensional manifold and should carry  fundamental homology
classes. According to a theorem of Freed and Uhlenbeck [13],
the moduli space $\CM^*$ of irreducible connections is
a smooth manifold with the actual dimension being equal
to the formal dimension for a generic choice of Riemann metric on $M$.
It is also known that there is no
reducible instanton for $b^+(M)>1$ [2].
For an odd $b^+(M) =1+2a$, the dimension of the moduli space is
even, $\hbox{\rm dim}\CM^* = 2d = 4k -3(1+a)$.
Donaldson also proved the
orientability of the moduli space [14].
The next and the most important
condition is to understand the compactness properties of $\CM^*$
so that one may define the fundamental homology classes.
In practice, the moduli space is rarely compact and this is the
main subtlety in defining the Donaldson invariants. However,
there is a natural compactification $\overline \CM^*$ of $\CM^*$ [15, 16].
One may extend the cohomology classes
$\m(\S_i)$ to the compactified space and try to evaluate their
cup products on $\overline \CM^*$.
For a large enough $k$
(or for the stable range $4k> 3b^+(M) + 3$), the compactified
space carries a fundamental homology class. In this way, we have a
well defined pairing. It is more convenient
to work in the Poincar\'{e} dual picture of \cupp.
Donaldson showed that one can arrange codimension $2$ cycles $V_{\S_i}$
--which represent the cohomology classes $\m(\S_i)$, on $\CM^*$ so that
the intersection
\eqn\zzz{
\CM^*\cap V_{\S_1}\cap\cdots\cap V_{\S_d}
}
is compact in the stable range. The Donaldson invariant is defined
by the intersection number of \zzz;
\eqn\dddod{
q_{k,M}\left(\S_1,\ldots,\S_d\right) = \#(\CM^*\cap
V_{\S_1}\cap\cdots\cap V_{\S_d}),
}
which is an invariant of the oriented diffeomorphism type of $M$.

\subsec{Witten's Interpretation and Generalization}
The TYMT in a special limit can be viewed as an integral representation
of the Donaldson invariants.
If we denote $\hW_{r}^{4-r}$ to be the restriction of $W_{r}{}^{4-r}$
on $\CB^*$ to $\CM^*$, the Donaldson invariants can be represented
by an integral of wedge products of the cohomology classes
$\hW_{r}{}^{4-r_i}\in H^{4-r_i}(\CM^*)$ over $\CM^*$,
\eqn\yyd{
\int_{\CM^*} \hW_{r_1}{}^{4-r_1}\wedge\cdots\wedge \hW_{r_k}{}^{4-r_k}\;,
}
which  vanishes unless $\sum_{i=1}^{k}(4-r_i) = \hbox{dim}(\CM^*)$.
Of course, we should have some concrete procedure
to obtain $\hW_r{}^{4-r}$ from $W_{r}{}^{4-r}$.

The basic idea of the TYMT is to introduce a fermionic symmetry $\dw$
and basic multiplet $(A,\P,\F)$ with the transformation law
analogous to \kac,
\eqn\kcc{
\dw A = \P\;,\qquad \dw \P = -d_{\!A}\F\;,\qquad \dw \F =0\;.
}
Geometrically, Witten's fermionic operator $\dw$ corresponds
to $\dh$. One can construct a $\dw$-invariant action after introducing
additional fields such that fixed point locus of the
fermionic symmetry ($\dw$-invariant configuration) is the moduli
space of ASD connections.
An appropriate action may be written as
\eqn\waltz{
S_W =\dw\biggl(\Fr{1}{h^2}\int_M
\tr \biggl[-\CX\wedge(H + F^+) + \bar\F (d_A*\P +\a\CY)]\biggr)\;,
}
where the self-dual two-forms $(\CX, H)$ (as well as the zero-forms
$(\bar\F,\CY)$) are analogous to the antighost multiplet for the
anti-self-duality constraint, $F^+=0$,
(and those for the horizontality $d_A *\P=0$ respectively)
in the usual BRST quantization,
\eqn\canon{\eqalign{
&\dw\bar\F=\CY\;,\cr
&\dw\CX=H\;,\cr
}\qquad\eqalign{
&\dw\CY=[\F,\bar\F]\;,\cr
& \dw H=[\F,\CX]\;.\cr
}}
This transformation law can be deduced from the
property that $\dw$ is nilpotent up to a gauge transformation
generated by $\F$ in the basic multiplet, i.e. $\dw^2 A =
-d_A\F$. These properties allow one to
interpret the TYMT as a BRST quantized version of an
underlying theory with topological symmetry [17--21].
We introduce a quantum number $U$, analogous to the
ghost number of BRST quantization, which assigns value $1$ to $\dw$
and $(0,1,2,-1,0,-2,-1)$ to $(A,\P,\F,\CX,H,\bar\F,\CY)$.
Note that the U numbers of the basic multiplet $(A,\P,\F)$
can be interpreted as the form degrees on $\CB^*$ in the universal
bundle formalism.

For $\a=0$, we find
\eqn\gigue{\eqalign{
S_W
=&\Fr{1}{h^2}\int_M \tr\biggl[-H\wedge(H+F^+) -\CX\wedge(d_A\P)^+
+\CX[\F,\CX] \cr
&+\CY d_A*\P-\bar\F(d_A*d_A\F+[\P,*\P])\biggr]\;.\cr
}}
We can integrate $H$ out by the Gaussian integral or by setting
$H=-\ha F^+$. The resulting action is $\dw$-invariant
if the $\CX$ transformation is modified to $\d\CX =
-\ha F^+$.
In the Witten's interpretation, the instanton cohomology groups
are realized by the zero-modes of the fermionic variables $(\CY,\P,\CX)$,
\eqn\ezero{
d_A\CY= 0\;,\qquad d_A^*\P= d_A^+\P= 0\;,\qquad d_A^{+*}\CX= 0\;.
}
For an example,
the $\CX$ and $\CY$ equations of motion
\eqn\whynot{
(d_A\P)^+ = 0\;,\qquad d_A*\P = 0\;,
}
show that a zero-mode of $\P$ represents a tangent vector
on the smooth part of instanton moduli space.
The number of the non-trivial solutions of \ezero\ for $(\CY,\P,\CX)$
are equal to $(h^0_A, h^1_A, h^2_A)$. Note that
the $U$ numbers of $\CY$, $\P$ and $\CX$  are $-1$, $1$ and $-1$ respectively.
Thus, the formal dimension of $\CM$ is just the number fermionic
zero-modes carrying $U=1$ minus the number of fermionic zero-modes
carrying $U=-1$.
If there are no $\CY$ and $\CX$ zero-modes,
$\CM$ is a smooth manifold with the actual dimension being equal to the
number of $\P$ zero-modes.
The $\bar\F$ integral leads to a delta function
constraint,
\eqn\jjj{
\F = -G_A[\P,*\P]\;,
}
which coincides with the universal bundle formalism \kaa.

Geometrically, observables of the TYMT correspond to
the de Rham cohomology classes $W_r{}^{4-r}$ on $\CB^*$ [11].
Correlation functions of observables can be formally written
as
\eqn\fixed{
<W_{r_1}{}^{4-r_1}\cdots W_{r_k}{}^{4-r_k}>
= \Fr{1}{\hbox{vol}(\CG)}\int (\CD X)
\exp(-S_W)W_{r_1}{}^{4-r_1} \cdots W_{r_k}{}^{4-r_k}\;.
}
Due to the fermionic symmetry, the path integral of the TYMT is
localized to an integral over the fixed point locus,
\eqn\partita{
\dw\CX = 0\;,\qquad \dw\P = 0\;,
}
which is the instanton moduli space and the space of $\F$
zero-modes, modulo gauge symmetry.
The localization of the path
integral to an integral over the fixed point locus is a general
property of any cohomological field theory [22].
Equivalently, the semiclassical limit, which
can be shown to be exact, of the path integral coincides with
the above localization. If there are no reducible connections, the
path integral reduces to an integral over the moduli space
$\CM^*$ of irreducible ASD connections.
Provided that there are no $\CY$ and $\CX$ zero-modes (that
is, $h^0_A$ and $h^2_A$ vanish everywhere), the path integral
(after integrating out non-zero modes) reduces to an integral
of wedge products of
closed differential forms on $\CM^*$,
\eqn\invariant{
<W_{r_1}{}^{4-r_1}\cdots W_{r_k}{}^{4-r_k}> = \int_{\CM^*}
\hW_{r_1}{}^{4-r_1}\wedge
\cdots\wedge \hW_{r_k}{}^{4-r_k}\;.
}
This is the integral representation \yyd\ of the Donaldson
invariants\fonote{See the four-steps (p.~381--383 in [1])
for the precise meanings.}.
The cohomology classes $\hW_{r_i}{}^{4-r_i}$ on $\CM^*$ can be
obtained from $W_{r_i}{}^{4-r_i}$ after replacing $F$ by its
instanton value, $\P$ by its zero-modes and $\F$ by the
zero-mode parts of its expectation value,
\eqn\kkg{
<\F> = -\int_M G_A[\P,*\P]\;.
}
Clearly, the integral \invariant\ will vanish unless the
integrand is a top form on $\CM^*$, $\sum_{i=1}^{k}(4-r_i) =
\hbox{dim}(\CM^*)$. Or, because the path integral measure have ghost
number anomaly due to the $\P$ zero-modes, Eq.\invariant\ will
vanish  unless we insert an appropriate set of observables with
net ghost number being equal to the number of $\P$ zero-modes.

One can see that the noncompactness
of the moduli space may make the topological interpretation
of Eq.\invariant\ problematic\fonote{This is briefly discussed in [23].}.
It is also unclear how
the compactification procedure  can be implanted
in the field theoretical interpretation.
However,
we should emphasis here that the viewpoint adopted in this
subsection is just a recipe to evaluate the correlation functions
under favorable situations.
It is important to note that
the energy-momentum tensor of the TYMT is $\dw$-exact,
\eqn\jjj{
T_{\a\b} = \dw \l_{\a\b}\;.
}
This leads that the correlation functions of the TYMT
give the Donaldson invariants, which are valid regardless of
whether instanton moduli space exists and what properties it
has [1]. In order to appreciate the real virtues of the TYMT,  however,
considerable progress in our understanding of quantum field theory
may be required.

\newsec{$N=2$ Topological Yang-Mills Theory}
Let $M$  be an $n$ complex dimensional compact K\"{a}hler
manifold endowed with a K\"{a}hler metric.  Picking a complex
structure on $M$ together with the K\"{a}hler metric, one can
determine K\"{a}hler form $\o$.  Let $\Omega^2$ be the space
of (real) two-forms on $M$.  Using the complex structure and the
K\"{a}hler form on $M$, one can decompose $\Omega^2$ as
\eqn\laa{
\Omega^2 = \Omega^2_+ \oplus \Omega^2_-\;,
}
where
\eqn\lab{\eqalign{
\Omega^2_+ &= \Omega^{2,0} + \Omega^{0,2} + \Omega^{1,1}_{\o}\;,\cr
\Omega^2_- &= \Omega^{1,1}_{\perp}\;.\cr
}}
We denote $\Omega^{1,1}_{\o}$ to be the space of $(1,1)$-forms
which is parallel to $\o$ and $\Omega^{1,1}_{\perp}$ to be the
orthogonal complement of $\Omega^{1,1}_\o$.
For an example, an element $\a$ of $\Omega^2_+$
can be written as $\a = \a^{2,0} + \a^{0,2} +
\a^{0}\o$ where $\a^{0,2}=\overline{\a^{2,0}}$ and
$\a^{0}$ is a real zero-form.
For $n=2$, the decompositions \laa\ and \lab\ are
identical to the decompositions of the space of two-forms
into the spaces of self-dual and anti-self two-forms.
Let $E$ be a vector bundle over $M$ with reduction of structure group
to $SU(2)$.
Let $\CA$ be the space of all connections
on $E$ and $\CG$ be the group of gauge transformations on $E$.
We can introduce a moduli space $\CM$
as the subspace of $\CB=\CA/\CG$ cut by the
following equations;
\eqn\lld{
F^{2,0}(A) =0\;,\qquad F^{0,2}(A)=0\;,\qquad \L F^{1,1}(A) =0\;,
}
where  $\L$ is the adjoint of the wedge multiplication
by $\o$
\eqn\ttr{
\L:\Omega^{p,q} \rightarrow \Omega^{p-1,q-1}\;.
}
In particular, the action of $\L$ on a $(1,1)$-form  measures the component
parallel to $\o$.

Behind the simple equation \lld, there is a beautiful
theorem of Donaldson-Uhlenbeck-Yau [24, 25], which
generalize a theorem of Narasimhan-Seshadri
[26] (see also Atiyah-Bott [27]), on the stable bundles:
An irreducible holomorphic bundle $\CE$ over a compact
K\"{a}hler manifold is stable if an only if there is an unique
unitary irreducible (Einstein-Hermitian) connection [28] on $\CE$ with
$\L F^{1,1}=c\times I_\CE$, where $c$ is a topological
invariant depending only on the cohomology classes
of $\o$ and $c_1(\CE)$ and $I_\CE$ is the identity endomorphism.
Note that each solution of Eq.\lld\ lies in the subspace
$\CA^{1,1}$, consisting of connections whose curvatures are type
$(1,1)$, of $\CA$. Thus, one can associate a holomorphic vector
bundle $\CE_A$ for each solution $A$. We can introduce the
moduli space of holomorphic vector bundles as a set of
isomorphism classes of holomorphic vector bundles.
The moduli space can be identified with the quotient space
$\CA^{1,1}/\CG^\BC$, where $\CG^\BC$ denotes the complexification of
$\CG$.  By the theorem of
Donaldson-Uhlenbeck-Yau, one can identify $\CM^*$
with the moduli space $\CM_M^s\subset \CA^{1,1}/\CG^\BC$
of stable holomorphic vector
bundles\fonote{See chapter $6$ in ref.\ [7] for details.}.
For instance, the moduli space of irreducible ASD $SU(2)$
connections is isomorphic to the moduli space of stable
$SL(2,\BC)$ bundles.  The moduli space of irreducible flat $SU(2)$
connections over a compact oriented Riemann surface $\S$
is also isomorphic to the moduli space of
stable $SL(2,\BC)$ bundles over $\S$.

Though the material in Sect.3.1 and in the later sections
 are generally valid (after slight modifications) on an arbitrary
dimensional compact K\"{a}hler manifold. We restrict our attention to
$n=1,2$ cases because we do not know the full details of the moduli
spaces otherwise.

\subsec{$N=2$ Algebra}
It is straightforward to obtain $N=2$ algebra from the basic
$\dw$ algebra using the natural complex structure on $\CB$
induced from $M$.  Picking a complex structure $J$ on $M$, one
can introduce a complex structure $J_{\!\CA}$ on $\CA$,
\eqn\cmme{
J_{\!\CA} \d A = J \d A\;,\qquad \d A \in T\CA\;,
}
by identifying $T^{1,0}\CA$ and $T^{0,1}\CA$ in $T\CA =
T^{1,0}\CA\oplus T^{0,1}\CA$ with $\adE$ valued $(1,0)$-forms
and $(0,1)$-forms on $M$ respectively
One can also introduce a K\"{a}hler structure $\tilde{\o}$
on $\CA$ given by
\eqn\kad{
\tilde\o = \Fr{1}{8\pi^2}\int_M \tr(\d A\wedge \d A)\wedge\o^{n-1}\;.
}
Note that the complex structure $J_\CA$ can descend to $\CA/\CG$
while the K\"{a}hler structure $\tilde\o$ does not in general.

Now we turn back to the universal bundle in Sect.2.1.
Using the complex structures $J$ and $J_{\CA}$,
we can decompose $A^{m,\ell}(M\times\CB^*)$ into
$A^{p,q,r,s}(M\times\CB^*)$, for $m=p+q$, $\ell=r+s$.
In terms of the form degree, one can decompose
$\CF^{1,1}$ as $\CF^{1,1} = \CF^{1,0,1,0} + \CF^{0,1,1,0} +\CF^{1,0,1,0}+
\CF^{1,0,0,1}$. The fact that $\CF^{1,1}=\dh A$ \kaa\
and the specific choice of the complex structure \cmme\
lead to $\CF^{1,1} = \CF^{1,0,1,0} +\CF^{0,1,0,1}$.
To see this more explicitly, we introduce the following decompositions of
the exterior covariant derivatives $d_A$ and $\dh$;
\eqn\sonata{
d_A = \Dp +\Dpp\;,\qquad \dh = \dh^\pr +\dh^\ppr\;,
}
where
\eqn\caa{\eqalign{
\Dp &:A^{p,q,r,s}(M\times\CB^*)\rightarrow A^{p+1,q,r,s}(M\times\CB^*)
\;,\cr
\Dpp&:A^{p,q,r,s}(M\times\CB^*)\rightarrow A^{p,q+1,r,s}(M\times\CB^*)
\;,\cr
\dh^\pr &:A^{p,q,r,s}(M\times\CB^*)\rightarrow A^{p,q,r+1,s}(M\times\CB^*)
\;,\cr
\dh^\ppr &:A^{p,q,r,s}(M\times\CB^*)\rightarrow A^{p,q,r,s+1}(M\times\CB^*)
\;.\cr
}}
Let $A = A^\pr +A^\ppr$ be the decomposition of $A$ into
the $(1,0)$ part $A^\pr$ and the $(0,1)$ part $A^\ppr$, then
we have $\dh A = \dh^\pr A^\pr + \dh^\ppr A^\ppr$.
The curvature two-form $\CF^{0,2}$ on $\CB^*$ can be decomposed
into its $(2,0)$, $(1,1)$, and $(0,2)$ components $\f$,
$\w$ and $\bar\f$. According to our choice of complex
structure\fonote{The Hodge star operator $*$ maps
a $r$-form on $M$ into $2n-r$-form
$$
*: \Omega^r(M)\rightarrow \Omega^{2n-r}(M)\;,
$$
and $*^2 =1$.
Note that $*$ maps a $(p,q)$-form on $M$ into $(n-q,n-p)$ form
$$
*: \Omega^{p,q}(M)\rightarrow \Omega^{n-q,n-p}(M)\;.
$$
{}From the definition of $\CF^{0,2}$ in Eq.\kaa, it follows that
$\CF^{0,2}$ is type $(1,1)$ on $\CB^*$.},
one can see
that the only non-vanishing component is $\w$.

Now we can deduce the $N=2$ algebra from the decompositions of the
basic $\dw$ algebra, Eq.\kac, according to the above structures.
Introducing the decompositions $\dw = \bs +\bbs$, $\P =\p +\bar\p$
and $\F =\w$, one can find that
\eqn\symphony{\eqalign{
&\bs A^\pr =\p\;,\cr
&\bbs A^\pr =0\;,\cr
&\bs A^\ppr =0\;,\cr
&\bbs A^\ppr =\bar\p\;,\cr
}\qquad\eqalign{
&\bs \p = 0\;,\cr
&\bbs\p = -\Dp\w\;,\cr
&\bs\bar\p = -\Dpp\w\;,\cr
&\bbs \bar\p = 0\;.\cr
}\qquad\eqalign{
\bbs\w=0\;,\cr
\bs\w=0\;.\cr
}}
We introduce another quantum number
$R$ which assigns values $1$ and $-1$ to $\bs$ and to $\bbs$
respectively.
Note that both $\bs$ and
$\bbs$ increase the $U$ number by $1$ since they are the decompositions
of $\dw$.
Geometrically, the fermionic operators $\bs$ and $\bbs$ correspond
to $\dh^\pr$ and $\dh^\ppr$ respectively. Then, a $(p,q,r,s)$-form
on $M\times \CB^*$ has the $U$ number $r+s$ and the $R$ number $r-s$.

One finds that $\bs$ and $\bbs$ are  nilpotent
and anti-commutative with each other
up to a gauge transformation generated by $\w$
\eqn\andante{
\bs^2 = 0\;,\qquad (\bs\bbs +\bbs\bs)A = -d_A\w\;,\qquad
\bbs^2  =0\;.
}
Geometrically, the operators $\bs$ and $\bbs$ can be interpreted as
the Dolbeault cohomology generators on $\CB^*$, provided that they
act on a gauge invariant quantity.
An arbitrary gauge invariant quantity $\a$ in the form $\a =\bs\bbs\b$
is invariant under the transformations
generated by both $\bs$ and $\bbs$. An $\bs$ and $\bbs$-closed form
$\a$ is automatically $\dw$-closed;
\eqn\bvva{
\bs\a =\bbs\a =0 \rightarrow (\bs +\bbs)\a =\dw\a =0\;.
}
On the other hand, the converse of the above relation,
\eqn\bvvb{
\dw\a = 0 \longrightarrow \bs\a = \bbs\a =0, \;\;\hbox{\it or}\;\;
\bs\a =0 \longleftrightarrow \bbs\a =0\;,
}
is  not generally valid, because the K\"{a}hler structure \kad\
on $\CA$ does not descend to $\CB$ in general. However, this is
not a real problem because the configuration space of the TYMT
is, due to the localization of the path integral, the moduli space
of ASD connection which has K\"{a}hler structure.
Furthermore, the failure
of Eq.\bvvb\ simply means the failure of the Hodge decomposition
theorem [29];
\eqn\bvvc{
H^r(\CB^*) \neq \sum_{p+q=r} H^{p,q}(\CB^*)\;.
}
Anyway, we can define the Dolbeault cohomology of $\CB^*$ by
the $\bbs$ operator, which satisfies the Hodge decomposition
after restriction to $\CM^*$. In Sect.4.1,
we will see that there is an important set of elements of
the Dolbeault cohomology of $\CB^*$ which are both $\bs$ and $\bbs$
closed.

To construct an action, we need to introduce more fields to impose
the anti-self-duality relative to the K\"{a}hler form,
\eqn\trio{
F^{2,0}=0\;,\qquad F^{0,2} =0\;,\qquad \L F^{1,1}=0\;.
}
{}From the relations \andante, one can establish the following
general transformation rules for an anti-ghost $\CJ$ and its
auxiliary fields $\CK,\bar\CK$;
\eqn\aux{\eqalign{
&\bs \CJ = \CK\;,\qquad      \bs\CK =\bs^2\CJ =0\;,\cr
&\bbs \CJ = \bar\CK\;,\qquad \bbs\bar\CK=\bbs^2\CJ =0\;,\cr
&(\bs\bbs +\bbs\bs)\CJ=\bs\bar\CK +\bbs\CK =[\w,\CJ]\;.\cr
}}
We  introduce a self-dual two-form $B$ with $(U,R)=(-2,0)$
which is the anti-ghost for the constraint \trio. We also introduce
the auxiliary fields $-\c$ and $\bar\c$ -- self-dual two forms carrying
$(U,R)=(-1,1)$  and $(U,R)=(-1,-1)$ respectively.
{}From \aux, we have
\eqn\mazuruka{\eqalign{
&\bs B = -\c\;,\qquad \bs\c = 0\;,\cr
&\bbs B =\bar\c\;,\quad  \qquad \!\bbs\bar\c = 0\;,\cr
&\bs\bar\c -\bbs\c =[\w,B]\;.\cr
}}
Due to the redundancy in the last relation, we can introduce one more
auxiliary field which can not be uniquely determined. We choose
$H\equiv \bs\bar\chi -\ha[\w,B]$,
a self-dual two form with $(U,R)=(0,0)$.  Then, one can find that
\eqn\symponieta{
\eqalign{
\bs\bar\c &= H +\ha[\w,B]\;,\cr
\bbs\c &=H-\ha[\w,B]\;,\cr
}\qquad
\eqalign{
\bs H &=\ha[\w,\c]\;,\cr
\bbs H &=\ha[\w,\bar\c]\;.\cr
}}
If we interpret $\c$ and $\bar\c$ as the
decompositions of $\CX$ according to the $R$ number, $\CX\equiv (\c
+\bar\c)/2$, our choice leads to
\eqn\interp{
(\bs +\bbs)\CX = H\;,\qquad (\bs+\bbs)H =[\w,\CX]\;,
}
which coincides with the $N=1$ algebra \canon.
But the anti-ghost $B$ and the commutator $\ha[\w,B]$ in \symponieta\ have
no counterparts in the $N=1$ algebra.
The $U$ and the $R$ numbers of the all fields introduced
in this subsection can be summarized by
\eqn\ghost{\matrix{
&\hbox{Fields}
&A^\pr&A^\ppr&\p&\bar\p&\w&B&\c&\bar\c&H\cr
&\hbox{$U$ Number}  &0&0&1&1&2&-2&-1&-1&0\cr
&\hbox{$R$ Number}  &0&0&1&-1&0&0&1&-1&0\cr
}\;.
}

\subsec{$N=2$ Actions}
Let $M$ be a compact K\"{a}hler surface with K\"{a}hler form $\o$.
The action for
the $N=2$ TYM on $M$ can be written  as
\eqn\vivace{
S = \bs\bbs\,{\bf B_T}\;,
}
which is a natural $N=2$ version of the Witten's action \waltz.
The unique choice of ${\bf B_T}$ with $(U,R)=(-2,0)$, so that
the action has $(U,R)=(0,0)$, is
\eqn\allegro{
{\bf B_T} =\int_M {\bf\CB_T}
= - \Fr{1}{h^2}\int_M \tr\biggl[B\wedge * F + \c\wedge *\bar\c
\biggr]\;.
}
Note that the self-dual two-forms $B,\c,\bar\c,H$
can be written as
\eqn\cae{\eqalign{
B &= B^{2,0} + B^{0,2} + B^{0}\o\;,\cr
H &= H^{2,0} + H^{0,2} + H^{0}\o\;,\cr
}\qquad\eqalign{
\c &= \c^{2,0} + \c^{0,2} + \c^{0}\o\;,\cr
\bar\c &= \bar\c^{2,0} + \bar\c^{0,2} + \bar\c^{0}\o\;,\cr
}}
which give rise to
\eqn\alle{\eqalign{
{\bf B_T} =
&-\Fr{1}{h^2}\int_M \tr\biggl[
  B^{2,0}\wedge * F^{0,2} +
  B^{0,2}\wedge * F^{2,0} +
  \c^{2,0}\wedge *\bar\c^{0,2}
  \cr
&+\c^{0,2}\wedge *\bar\c^{2,0}
  +\biggl(B^{0} f + \c^{0}\bar\c^{0}\biggr)\o^2
  \biggr]\;,
  \cr
}}
where $f \equiv\ha \hat F = \ha\L F^{1,1}$.
{}From \symphony, \mazuruka, \symponieta, \vivace\ and \alle, we find that
\eqn\moderato{\eqalign{
S =
&\Fr{1}{h^2}\!\int_M\!\!\! \tr\biggl[
 - H^{2,0}\wedge*\left(H^{0,2}+F^{0,2}\right)
 - H^{0,2}\wedge*\left(H^{2,0}+F^{2,0}\right)
 -\c^{2,0}\wedge *\Dpp\bar\p
 \phantom{\biggl]}
 \cr
&-\bar\c^{0,2}\wedge *\Dp\p
 -[\w,\c^{2,0}]\wedge *\bar\c^{0,2}
 -[\w,\c^{0,2}]\wedge *\bar\c^{2,0}
 +\ha[\w,B^{2,0}]\wedge * F^{0,2}
 \phantom{\biggl]}
 \cr
&-\ha[\w,B^{0,2}]\wedge * F^{2,0}
 +\ha[\w,B^{2,0}]\wedge *[\w,B^{0,2}]
 -\biggl( H^{0}\left(H^{0}+f\right)
 \phantom{\biggl]}
 \cr
&+\ha\bar\c^{0}\L \Dpp\p
 +\ha\c^{0}\L \Dp\bar\p
 +[\w,\c^{0}]\bar\c^{0}
 -\fr{1}{4}[\w,B^0]^2
 \phantom{\biggl]}
 \cr
&-\ha B^{0}\L\left(\Dp \Dpp\w +[\p,\bar\p] +[\w,f]\right)
\biggr)\o^2\biggr]\;,
 \cr
}}
where we have used $\Dpp\Dpp\w = [F^{0,2},\w]$.
We can integrate out $H^{2,0}$, $H^{0,2}$ and $H^{0}$ from the
action by setting $H =-\ha F^+$ or by the Gaussian integral, which
leads to a modified transformation law
\eqn\jane{
\eqalign{
&\bs\bar\c^{2,0} = -\ha F^{2,0}+\ha[\w,B^{2,0}]\;,\cr
&\bs\bar\c^{0,2} = -\ha F^{0,2}+\ha[\w,B^{0,2}]\;,\cr
&\bs\bar\c^{0} = -\ha f+\ha[\w,B^{0}]\;,\cr
}\qquad\eqalign{
&\bbs\c^{2,0} = -\ha F^{2,0}-\ha[\w,B^{2,0}]\;,\cr
&\bbs\c^{0,2} = -\ha F^{0,2}-\ha[\w,B^{0,2}]\;,\cr
&\bbs\c^{0} = -\ha f-\ha[\w,B^{0}] \;.\cr
}}
One can see that the locus of $\bs$ and $\bbs$ fixed points in
the above transformations is precisely the space of ASD connections.
Due to the localization principle of the cohomological field
theories [22], the path integral reduces to the integral over the locus
of the $\bs$ and $\bbs$ fixed points, $\bs\c =\bbs\bar\c
=0,\;\bs\bar\p=\bbs\p =0$, which is the instanton moduli space
with the space of $\w$ zero-modes ($d_A\w=0$).

Note that the terms $\tr\left(-\ha[\w,B]\wedge *F
+ \fr{1}{4}[\w,B]\wedge *[\w,B]\right)$ in the action can be dropped
without changing the theory, since: i) they do not change the fixed
point locus; ii) the first term vanishes at the fixed point locus;
iii) the $B^{2,0}$ and $B^{0,2}$ equations of motion are trivial
algebraic ones; iv) they do not contribute to the $B^0$ equation of
motion, provided that the theory localizes to the fixed point locus.
The terms containing the commutator, $[\w,B]$, are trivial  because
the commutator is originated from the gauge degree of freedom in the
anti-ghost multiplet, as we can see
from Eq.\mazuruka.
However, the another type of gauge degree of freedom appearing in the
basic transformation law  \symphony\ should be
maintained since we are dealing with the
equivariant cohomology.
In our action functional, it gives a very
important term, $\ha\tr B^{0}\L\left(\Dp \Dpp\w +[\p,\bar\p]\right)\o^2$,
which leads to the $B^0$ equation of motion,
\eqn\curva{
i\Dpp^*\Dpp\w +\L[\p,\bar\p]=0\;,
}
where we have used the K\"{a}hler identities;
\eqn\cai{
\Dp^* = i[\L,\Dpp]\;,\qquad
\Dpp^* = -i[\L,\Dp]\;.
}

Now we can choose a simpler transformation law
for the anti-ghost multiplet by dropping
$[\w,B]$ term in Eq.\mazuruka,
\eqn\simpler{
\eqalign{
&\bs B = -\c\;,\cr
&\bbs B =\bar\c\;,\cr
}\qquad\eqalign{
&\bs\c = 0\;,\cr
&\bbs\bar\c = 0\;,\cr
}\qquad\eqalign{
&\bs\bar\c = H\;,\cr
&\bbs\c =H\;,\cr
}\qquad
\eqalign{
&\bs H =0\;,\cr
&\bbs H =0\;.\cr
}}
The corresponding action is
\eqn\varactiona{\eqalign{
S =&\Fr{1}{h^2}\!\int_M\! \tr\biggl[
    - H^{2,0}\wedge*\left(H^{0,2}+F^{0,2}\right)
    - H^{0,2}\wedge*\left(H^{2,0}+F^{2,0}\right)
    \cr
   &+B^{2,0}\wedge *[F^{2,0},\w]
    -\c^{2,0}\wedge *\Dpp\bar\p
    -\bar\c^{0,2}\wedge *\Dp\p
    \phantom{\biggl]}
    \cr
   &-\biggl( H^{0}\left(H^{0}+f\right)
    +\ha\bar\c^{0}\L \Dpp\p
    +\ha\c^{0}\L \Dp\bar\p
    \cr
   &-\ha B^{0}\L\left(\Dp \Dpp\w +[\p,\bar\p]\right)\biggr)\o^2
    \biggr]\;.
    \cr
}}
One can easily check that this action is equivalent to the original one.
We can also consider the simplest possible action of the $N=2$ TYMT
by discarding $\tr \c\wedge *\bar\c$ term from Eq.\allegro\
and using the transformation law \simpler,
\eqn\varactionb{\eqalign{
S =
&\Fr{1}{h^2}\!\int_M\! \tr\biggl[
 - H^{2,0}\wedge*F^{0,2}
 - H^{0,2}\wedge*F^{2,0}
 +B^{2,0}\wedge *[F^{2,0},\w]
 -\c^{2,0}\wedge *\Dpp\bar\p
 \cr
&-\bar\c^{0,2}\wedge *\Dp\p
 -\biggl( H^{0}f
 +\ha\bar\c^{0}\L \Dpp\p +\ha\c^{0}\L \Dp\bar\p
 \cr
&-\ha B^{0}\L\left(\Dp \Dpp\w +[\p,\bar\p]\right)\biggr)\o^2
 \biggr]\;.
 \cr
}}
The localization of this action to the moduli space of ASD
connections is provided by the delta function gauges instead of
the fixed point argument. It is a typical property of general cohomological
field theory that there can be different realizations of the same theory.

\subsec{Fermionic Zero-Modes}
Now we will discuss about the geometrical meaning of the
zero-modes of the various fermionic fields
$(\p,\bar\p,\c^{2,0},\bar\c^{0,2},\c,\bar\c)$.
Let $\CE_A$ be a holomorphic structure induced by an ASD connection
$A$. Let $\hbox{End}^0(\CE_A)$ be the trace-free enomorphism bundle
of $\CE_A$.

The
$\c^{2,0}$, $\bar\c^{0,2}$, $\c^{0}$, and $\bar\c^{0}$
equations of motion (up to gauge transformations) are
\eqn\largoo{
\Dpp\bar\p=0\;,\qquad \Dp\p =0\;,\qquad \L \Dpp\p =0\;,\qquad
\L \Dp\bar\p =0\;.
}
Using the K\"{a}hler identities \cai,
one finds that a zero-mode of $\bar\p$ becomes
an element of $(0,1)$-{\it th} twisted Dolbeault cohomology group
of $\hbox{End}^0(\CE_A)$,
\eqn\vva{
{\bf H}^{0,1} = \{\Dpp\bar\p=0\quad \hbox{and}\quad \Dpp^*\bar\p =0\}\;.
}
The geometrical meaning of the zero-modes of $\p$ and
$\bar\p$, which are the non-trivial solutions
of Eq.\largoo, can be easily understood
as follows ([3] and Chap.7.2 of \ [28]);
for a given ASD connection $A =A^\pr + A^\ppr$,
\eqn\caf{
F^{2,0}(A^\pr) =0\;,\qquad F^{0,2}(A^\ppr) =0\;,\qquad \L
F^{1,1}(A^\pr,A^\ppr) =0\;,
}
an infinitely close instanton solution $A +\d A$ (after
linearizations) should satisfy
\eqn\cag{\eqalign{
F^{2,0}(A^\pr +\d A^\pr) &\sim \Dp \d A^\pr =0\;,\cr
F^{0,2}(A^\ppr +\d A^\ppr) &\sim \Dpp \d A^\ppr =0\;,\cr
\L F^{1,1}(A^\pr +\d A^\pr,A^\ppr +\d A^\ppr) &\sim
\L(\Dpp \d A^\pr +\Dp\d A^\ppr)=0\;.\cr
}}
In addition, we require $\d A$ to be orthogonal to the pure gauge
variation,
\eqn\cah{
d_A^*\d A = 0\;,
}
so that a $\d A$ satisfying \cag\ and \cah\ represents a
tangent vectors at $A$ on $\CM$.
Using the K\"{a}hler identities \kad,
one can rewrite \cah\ as
\eqn\caj{
\L(\Dpp\d A^\pr - \Dp\d A^\ppr)=0\;.
}
Combining \cag\ and \caj, we can see that the zero-modes of $\p$ and
$\bar\p$ are holomorphic and anti-holomorphic tangent vectors,
respectively, on the smooth part of $\CM$. Thus, we have
\eqn\vvb{
H_A^1 \approx {\bf H}^{0,1}\;,\qquad h^1_A = 2{\bf h}^{0,1}\;,
}
where ${\bf h}^{0,q}\equiv \hbox{dim}_\BC {\bf H}^{0,q}(\hbox{End}^0(\CE_A))$.

One can also find that the zero-modes of
$\c^{2,0},\bar\c^{0,2},\c^{0}$ and $\bar\c^{0}$
 satisfy the following
equations (the $\p$ and $\bar\p$ equations of motion);
\eqn\cak{
\eqalign{
&\Dp^*\c^{2,0} +i \Dp\c^{0}=0\;,\cr
&\Dpp^*\bar\c^{0,2}-i\Dpp\bar\c^{0}=0\;,\cr
}}
with obvious relations $\Dp\c^{2,0}=\Dpp\bar\c^{0,2}=0$ coming from the
dimensional reasoning.
Note that an self-dual two form $\a$ satisfies [3]
\eqn\xdx{
d_A^{*+}\a =\Dp^* \a^{2,0} + \Dpp^*\a^{0,2} +
i(\Dp\a^{0} - \Dpp\a^0)\;.
}
If $A$ is an ASD connection\fonote{Similar equation is also valid
for an EH connection on a compact K\"{a}hler manifold.}, we have
\eqn\evo{
\left|d_A^{*+}\a\right|^2=
\left|\Dp^*\a^{2,0}\right|^2
+\left|\Dpp^*\a^{0,2}\right|^2
+\left|d_A\a^{0}\right|^2\;.
}
Thus, Eq.\cak\ reduces to
\eqn\ewq{
\Dp^*\c^{2,0}=\Dp \c^{0} =0\;,\qquad
\Dpp^*\bar\c^{0,2}=\Dpp \bar\c^{0} =0\;.
}
In
particular, a zero-mode of $\bar\c^{0,2}$ is an element of
the $(0,2)$-{\it th} twisted Dolbeault cohomology
\eqn\vvb{
{\bf H}^{0,2}
= \{\Dpp\bar\c^{0,2}=0\quad \hbox{and}\quad \Dpp^*\bar\c^{0,2}=0\}\;,
}
and a zero-mode of $\bar\c^0$ is an element of $H_A^{0}$.

An important point
is that the zero-mode of the fermionic field $\p$ and
those of $\c$ are always accompanied by their counterparts,
i.e. the zero-modes of $\bar\p$ and $\bar\c$ respectively.
Thus, the net violation of the $R$ number by the fermionic zero-modes
is always zero,
while the net violation of the $U$ number by the fermionic zero-modes
being equal to $2({\bf h}^{0,1} - h_A^{0} - {\bf h}^{0,2})$. That is,
the half of the net $U$ number violation is identical to the number of
$\bar\p$ zero-modes minus the number of $\bar\c^{0}$ zero-modes
minus the number of $\bar\c^{0,2}$ zero-modes.

The various fermionic zero-modes of the $N=2$
TYMT naturally realize the cohomology group of
the Atiyah-Hitchin-Singer and Itoh's instanton complex [3, 12]\fonote{
The generalization to the moduli space of EH connections on
a compact K\"{a}hler manifold was studied by Kim [30]. See also Chap.7
of [28].},
\eqn\eccc{

\def\mapright#1{\!\!\!\smash{
    \mathop{\longrightarrow}\limits^{#1}}\!\!\!}
\def\mapdown#1{\Big\downarrow
  \rlap{$\vcenter{\hbox{$\scriptstyle#1$}}$}}
\matrix{
0&\mapright{}&\O^0(\adE)&\mapright{d_{\!A}}&\O^1(\adE)
&\mapright{d_{\!A}^+}&\O^2_+(\adE)
&\mapright{}&\!0\cr
&&\mapdown{} &&\mapdown{} &         &\mapdown{} & \cr
0&\mapright{}&\O^{0,0}(\hbox{End}^{0}(\CE_A))
&\mapright\Dpp&\O^{0,1}(\hbox{End}^{0}(\CE_A))
&\mapright\Dpp&\O^{0,2}(\hbox{End}^{0}(\CE_A))
&\mapright{}&0\;.\cr
}
}
The first elliptic complex is the original Atiyah-Hitchin-Singer's
instanton complex (Eq.\eic) and its cohomology
$H^q_A$ (with $h^q_A\equiv \hbox{dim}_\BR H^q_A$))
was given by Eq.\ecoh.
The second one is the twisted Dolbeault complex of
$\hbox{End}^{0}(\CE_A)$
and its cohomology group was denoted by
${\bf H}^{0,q}$ (with ${\bf h}^{0,q}\equiv \hbox{dim}_\BC {\bf
H}^{0,2})$).
Then (Theorem $7.2.21$ and Eq.(7.2.29) in [28]), we have
\eqn\thenr{
\eqalign{
&H^0_A\otimes C \approx {\bf H}^{0,0}\;,\cr
&H^1_A \approx {\bf H}^{0,1}\;,\cr
&H^2_A \approx {\bf H}^{0,2}\oplus H^0_A\;,\cr
}\qquad
\eqalign{
&h^0_A ={\bf h}^{0,0}\;,\cr
&h^1_A =2{\bf h}^{0,1}\;,\cr
&h^2_A =2{\bf h}^{0,2} +h^0_A\;,\cr
}
}
In particular,
\eqn\oiyhk{
\hbox{dim}_\BR(\CM) = h^1_A - h^0_A - h^2_A =
2({\bf h}^{0,1} - h_A^{0} - {\bf h}^{0,2})\;.
}

Thus the formal (real) dimension of the moduli space is
identical to the net violation of the $U$ number by the fermionic zero-modes,
i.e. the $U$ number anomaly.
Note that there are no zero-modes except $\p$ and
$\bar\p$ pairs if and only if the cohomology group $H^2_A$ is trivial.
Then, the moduli space is a smooth K\"{a}hler manifold with
complex dimension being equal to the number of $\bar\p$ zero-modes.
For example, if we consider a simply connected
compact K\"{a}hler surface with a strictly positive
geometric genus $p_g(M) > 0$. From the the formula $b^+(M) = 1 + 2p_g(M)$,
we have $b^+(M)\geq 3$ and, then, it is known that there are no reducible
instantons for a generic choice of metric.  It is also known that for a
large enough integer $k$, the second cohomology group vanishes
at least over a dense subset of $\CM^*$,
or for a dense set of stable bundles $\CE_A$ the cohomology group
${\bf H}^{0,2}$ vanishes [2].
Then, the generic part in the instanton
moduli space is a smooth K\"{a}hler manifold with complex dimension,
$\hbox{dim}_\BC(\CM)= 4k -3(1+p_g(M))$.

Finally, we note that the zero-modes of bosonic
variables $A,\w,B^0$ do not contribute to the $U$ number violation
in the path integral measure. This is because the numbers of $\w$ and
$B^0$ zero-modes are the same, each arising from a reducible connection,
while they carry the same $U$ numbers up to sign. Thus, the $U$ number
anomaly is an obstruction to having a well defined path integral measure,
which should be absorbed by inserting appropriate observables [1].

\newsec{Observables and Correlation Functions}
\subsec{Observables}
{}From an obvious  $\bs$ and $\bbs$ invariant
$\CW_{0,0}{}^{2,2}=\ha\tr\w^2$, which are gauge invariant and
metric independent\fonote{ Our particular choice is that our main
interest in this paper is the $SU(2)$ invariants[1]. We can
consider a general invariant polynomial in $\w$ obeying all the
criterions for a topological observable. And, the basic results
discussed below can be applied without modifications.  },
one can
find the following topological descent equations;
\eqn\rondo{
\bs \CW_{p,q}{}^{r,s}+
\bbs \CW_{p,q}{}^{r+1,s-1}+
\rd \CW_{p-1,q}{}^{r+1,s}+
\bar\rd \CW_{p,q-1}{}^{r+1,s}=0\;,
}
where we generally denotes $\CW_{\a,\b}{}^{\r,\d}$  as an
${\a,\b}$-form on $M$ and $(\r,\d)$ form of $\CB$ and
all the indicies are positive (that is, $\CW_{\a,\b}{}^{\r,\d}$
with negative indicies vanish in the above equation),
$\a+\b+\r+\d = 4$ and $\a,\b,\r,\d=0,1,2$.
Explicitly,
\eqn\moonlight{
\eqalign{
\CW_{0,0}{}^{2,2}&= \ha\tr(\w^2)\;,\cr
\CW_{1,0}{}^{2,1}&=\tr(\p\w)\;,\cr
\CW_{0,1}{}^{1,2}&=\tr(\bar\p\w)\;,\cr
\CW_{2,0}{}^{2,0}&=\ha\tr(\p\wedge\p)\;,\cr
\CW_{2,0}{}^{1,1}&=\tr(\w F^{2,0} )\;,\cr
\CW_{0,2}{}^{0,2}&=\ha\tr(\bar\p\wedge\bar\p)\;,\cr
\CW_{0,2}{}^{1,1}&=\tr(\w F^{0,2})\;,\cr
}\qquad\eqalign{
\CW_{1,1}{}^{1,1}&=\tr(\w F^{1,1}+\p\wedge\bar\p)\;,\cr
\CW_{2,1}{}^{0,1}&=\tr(\bar\p\wedge F^{2,0})\;,\cr
\CW_{2,1}{}^{1,0}&=\tr(\p\wedge F^{1,1})\;,\cr
\CW_{1,2}{}^{0,1}&=\tr(\bar\p\wedge F^{1,1})\;,\cr
\CW_{1,2}{}^{1,0}&=\tr(\p\wedge F^{0,2})\;,\cr
\CW_{2,2}{}^{0,0}&=\tr(F^{0,2}\wedge F^{2,0}+\ha F^{1,1}\wedge F^{1,1})\;,\cr
}}
and all the other components vanish.
Note that the Eq.\rondo\ and \moonlight\ can be combined into
a single identity,
\eqn\single{
(\rd +\bar\rd +\bs+\bbs)\Ha\tr (F^{2,0}+F^{0,2}+F^{1,1}
+\p+\bar\p+\w)^2=0\;.
}

Let
\eqn\dab{
W_{\a,\b}{}^{\r,\d}\equiv
\Fr{1}{4\pi^2}\int_M \CW_{\a,\b}{}^{\r,\d}\wedge
O_{2-\a,2-\b}\;,
}
where $O_{2-\a,2-\b}\in H^{2-\a,2-\b}(M)$
denotes a harmonic $(2-\a,2-\b)$-form on $M$.
Equation \rondo\ gives
\eqn\pathetique{
\bs  W_{p,q}{}^{r,s} + \bbs W_{p,q}{}^{r+1,s-1}=0\;,
}
where we have used the condition that $M$ is a compact K\"{a}hler
surface.

Because $\CB$, in general, do not have K\"{a}hler structure,
Eq.\pathetique\ does not imply that every $W_{\a,\b}{}^{\r,\d}$ is
both $\bs$ and $\bbs$ closed. For an example,
\eqn\taaa{
\bs W_{2,1}{}^{0,1} =-\Fr{1}{4\pi^2}\int_M \tr(\Dpp\w\wedge F^{2,0}
+\bar\p\wedge\Dp\p)\wedge O_{0,1}\neq 0\;.
}
If we restrict $W_{p,q}{}^{r,s}$ to
$\CM^*$, all of them are both $\bs$
and $\bbs$ closed due to the K\"{a}hler structure on $\CM^*$.
In quantum field theory, however, it is unnatural
to use an quantity which is invariant only on-shell as an observable, though
it is not entirely impossible.
Fortunately, we can find a set of well defined topological observables,
\eqn\taab{\eqalign{
W_{0,0}{}^{2,2} &=\Fr{1}{8\pi^2}\int_M \tr(\w^2)\Fr{\o^2}{2!}\;,\cr
W_{1,0}{}^{2,1} &=\Fr{1}{4\pi^2}\int_M \tr(\p\w)\wedge O_{1,2}\;,\cr
W_{0,1}{}^{1,2} &=\Fr{1}{4\pi^2}\int_M \tr(\p\w)\wedge O_{2,1}\;,\cr
W_{1,1}{}^{1,1} &=\Fr{1}{4\pi^2}\int_M \tr(\w F^{1,1}+\p\wedge\bar\p)
\wedge O_{1,1}\;,\cr
W_{2,2}{}^{0,0} &=\Fr{1}{4\pi^2}\int_M
\tr(F^{0,2}\wedge F^{2,0}+\ha F^{1,1}\wedge F^{1,1})\;.\cr
}
}
Using the Bianchi identity, $d_A F_A =0$, and integration by parts,
one can see that they are both $\bs$ and $\bbs$ closed.
Geometrically, an observable ${W}_{p,q}{}^{r,s}$
is an element of $(r,s)$-{\it th} Dolbeault cohomology group on $\CB^*$,
$W_{p,q}{}^{r,s}\in H^{r,s}(\CB^*)$, which depends only on
the cohomology class of $O_{2-p,2-q}$ in $M$.
Finally, we note an interesting descent equations
among $\CW_{0,0}{}^{2,2}, \CW_{1,1}{}^{1,1}$ and $\CW_{2,2}{}^{0,0}$,
\eqn\taac{\eqalign{
\bs\bbs\;\tr(\w F^{1,1}+\p\wedge\bar\p) &= -\rd\bar\rd\;\ha tr(\w^2)\;,\cr
\bs\bbs\;\tr(F^{0,2}\wedge F^{2,0}+\ha F^{1,1}\wedge F^{1,1})
&=-\rd\bar\rd\;\tr(\w F^{1,1}+\p\wedge\bar\p)\;.\cr
}}

\subsec{Correlation Functions}
In principle, we can give formulas for the Donaldson invariants on an
arbitrary compact K\"{a}hler surfaces for an arbitrary compact group.
Following the manipulations developed by Witten, we can formally
show that the correlation functions of the $N=2$ TYMT are topological
invariants in general circumstances.  It is crucial to prove
that the energy-momentum tensor $T_{\a\bar \b}$ defined by the
variation of the action under an infinitesimal change of the
K\"{a}hler metric $g^{\a\bar\b}\rightarrow g^{\a\bar\b} +\d
g^{\a\bar\b}$,
\eqn\wwa{
\d S = \Ha\int_M \sqrt{g}\d g^{\a\bar\b}\, T_{\a\bar\b}\;,
}
is a $\bs$ and $\bbs$ exact form
\eqn\wwb{
T_{\a\bar\b} = \bs\bbs \l_{\a\bar\b}\;.
}
The above relation is an immediate consequence of \vivace,
if the variation operator
$\d/\d g^{\a\bar\b}$ commutes to
both $\bs$ and $\bbs$ off shell.
The only subtlety here is that the various anti-ghost and
auxiliary fields $B,\c,\bar\c,H$ are subject to the self-duality
constraint \cae\ which must be preserved after the deformation
of the metric [1]. Clearly, the variations of
$B,\c,\bar\c,H$ come only from their $(1,1)$ components, which should
be remained parallel to the new K\"{a}hler form associated with the
deformed K\"{a}hler metric [6]. Thus, an arbitrary change $\d
g^{\a\bar\b}$ in the K\"{a}hler metric must be accompanied by
\eqn\wwc{
g^{a\bar\b}\d H_{\a\bar\b} = -H_{\a\bar\b}\d g^{\a\bar\b}\;.
}
We have the same relations for $\c_{\a\bar\b}$, $\bar\c_{\a\bar\b}$
and $B_{\a\bar\b}$.  One can easily see that $\d/\d
g^{\a\bar\b}$ commutes to $\bs$ and $\bbs$ off shell.
Consequently, we have
\eqn\wwd{
T^{\a\bar\b} = \bs\bbs \l_{\a\bar\b}\;,\qquad
\l_{\a\bar\b} = \Fr{2}{\sqrt{g}}\Fr{\d}{\d
g^{\a\bar\b}}{\bf\CB_T}\;.
}

Now it is an easy exercise to show that the correlation
functions of the observables are topological invariants by
following the essentially same manipulation as given in Sect.3
of [1].  Here, we will only note about the ghost number
anomaly. Recall that the action \moderato\ has two global
symmetries generated by $U$ and $R$ charges at the classical level.
If the formal dimension of the moduli space is non-zero,
we have an $U$ number anomaly
because the net $U$ ghost number violation $\triangle U$
coincides with the formal dimension.
Let $\left\{W_{p_i,q_i}{}^{r_i,s_i}\right\}$ be the general
topological observables which can be constructed from
$\CW_{0,0}{}^{2k,2k} =\ha\tr\w^{2k}$.
Then, a correlation function
\eqn\apata{
\biggl<\prod_{i=1}^{\ell} W_{p_i,q_i}{}^{r_i,s_i}\biggr> =
\Fr{1}{\hbox{vol}(\CG)}\int (\CD X)\exp(-S)\cdot
\prod_{i=1}^{\ell}
W_{p_i,q_i}{}^{r_i,s_i}\;,
}
vanishes unless $\prod_{i=1}^{\ell}W_{p_i,q_i}{}^{r_i,s_i}$ carries
the $U$ number $\triangle U$ and the $R$ number zero.
Thus, we have the following superselection rule;
\eqn\iii{
\hbox{dim}_\BR(\CM) = \triangle U = \sum_{i=1}^{\ell}(r_i+s_i)\;,
\qquad \sum_{i=1}^{\ell}(r_i -  s_i) =0\;,
}
which is equivalent to the condition
$\hbox{dim}_\BC(\CM) = \sum_{i=1}^\ell r_i=\sum_{i=1}^\ell s_i$.

\subsec{Differential Forms on Moduli Space}
Throughout this sub-section, we assume that there are the
zero-modes of $(\psi,\bar\psi)$ pairs only.
In this case, the path integral localizes to the moduli space $\CM^*$ of
irreducible ASD connections which is a finite dimensional
smooth K\"{a}hler manifold. Let $\hbox{\rm dim}_\BC(\CM^*)\equiv d$,
the number of the $(\psi,\bar\psi)$ zero-modes pairs.
The basic result of the $N=1$ TYMT is that the expectation value
of some products of observables reduces to an integral of wedge products of
the de Rham cohomology classes on $\CM^*$ over $\CM^*$ [1].  In
the $N=2$ TYMT theory, the path integral will reduce to an integral of
wedge products of the Dolbeault cohomology classes on $\CM^*$.

We can obtain an element $\hW_{p,q}{}^{r,s}\in H^{r,s}(\CM^*)$
of Dolbeault cohomology group on $\CM^*$ from
$W_{p,q}{}^{r,s}$  after restricting to $\CM^*$.
Clearly, this restriction is provided by the localization
of the path integral. In the exact semiclassical limit, a
correlation function
\eqn\appassionata{
\biggl<\prod_{i=1}^{n} W_{p_i,q_i}{}^{r_i,s_i}\biggr> =
\Fr{1}{\hbox{vol}(\CG)}\int (\CD X)\exp(-S)\cdot \prod_{i=1}^{n}
W_{p_i,q_i}{}^{r_i,s_i}
}
reduces to an integral of wedge products
of the elements of $H^{r_i,s_i}(\CM^*)$,
\eqn\jack{
\biggl<\prod_{i=1}^{n} W_{p_i,q_i}{}^{r_i,s_i}\biggr>
= \int_{\CM^*} \hW_{p_1,q_1}{}^{r_1,s_1}\wedge\cdots\wedge
\hW_{p_n,q_n}{}^{r_n,s_n}\;,
}
after integrating all non-zero modes out.  Clearly, the
correlation function \jack\ vanishes unless  the integrand is a
top form, a $(d,d)$-form, on $\CM^*$,
\eqn\rock{
\sum_{i=1}^{n}(r_i,s_i) = (d,d)\;.
}
This condition coincides with the superselection rule
due to the ghost number anomalies
\eqn\rock{
\triangle R =\sum_{i=1}^{n}(r_i+s_i) = 2d\;,\qquad
\triangle U =\sum_{i=1}^{n}(r_i-s_i) = 0\;.
}
Explicitly, $\hW_{p,q}{}^{r,s}$ can be
obtained from $W_{p,q}{}^{r,s}$ by replacing
\eqn\lark{\eqalign{
&\bullet\hbox{$F$ with its instanton value}\;,\cr
&\bullet\hbox{$\p$ and $\bar\p$ with their zero-modes}\;,\cr
&\bullet\hbox{$\w$ with the zero-mode parts of $<\w>
=\int_M d\m\, \Fr{i}{\Dpp^*\Dpp}\L[\p,\bar\p]$}\;,\cr
}}
where the last relation is resulted from the $B^{0}$ equation of
motion \curva.

We note that the smooth part of the moduli space of ASD connections
is a K\"{a}hler manifold. This was
first proved by Itoh using a direct but difficult calculation [3].
His result is a generalization of a theorem of Atiyah and Bott
on the K\"{a}hler structure of the moduli space of flat connections
on Riemann surfaces [27]. It was futher generalized to the moduli space
of EH connections on an arbitrary dimensional compact K\"{a}hler
manifold by Kobayashi. Let $M$ be an $n$ complex dimensional
compact K\"{a}hler manifold.
The K\"{a}hler structure on the smooth part of
the moduli space of EH connections  can be proved very
compactly  by the symplectic (Mardsen-Weinstein) reduction
(Chap.7.6 in [28] and Chap.6.5.1--3 in [7]).
First, we restrict
$\CA$ to the subspace $\CA^{1,1}\subset \CA$ consisting of connections
having curvature of type $(1,1)$. The subspace $\CA^{1,1}$ is
preserved by the action of $\CG$ (as well as by the action of $\CG^\BC$)
and its smooth part has the K\"{a}hler structure given by the restriction
of $\kad$. Let $Lie(\CG)$ be the Lie algebra of $\CG$,
which can be identified with the space of $\adE$-valued zero-forms.
Then, we have a moment map
$\frak{m}:\CA^{1,1}\rightarrow \O^0(\adE)^*$
\eqn\taad{
\frak{m}(A) = -\Fr{1}{4\pi^2}F^{1,1}_{\!A}\wedge\o^{n-1}
= -\Fr{1}{4\pi^2}f\o^n\;,
}
where $\O^0(\adE)^*=\O^{2n}(\adE)$ denotes the dual of $\O^0(\adE)$.
The reduced phase space (or the symplectic quotient)
$\frak{m}^{-1}(0)/\CG$
can be identified with the moduli space $\CM$ of EH connections.
Then the smooth part of the reduced phase space has
K\"{a}hler structure descended from $\CA^{1,1}$ by
the reduction theorem of Mardsden and Weinstein.
We will also denote the K\"{a}hler structure on $\CM$ by
$\tilde\o$. One can also consider
the semi-stable set $\CA^{1,1}_{ss}\subset \CA^{1,1}$
and its ordinary (complex) quotient space
$\CA^{1,1}_{ss}/\CG^\BC \equiv \CM^{ss}_M$ which is isomorphic
to  the symplectic quotient $\frak{m}^{-1}(0)/\CG=\CM$.
Then, the smooth part of $\CM^{ss}_M$ (which is identical to the smooth
part of $\CM^s_M$) has the K\"{a}hler structure.

Let $M$ be a  compact
K\"{a}hler surface. The second cohomology $H^2(M)$
is always non-trivial due to the cohomology class represented
by the K\"{a}hler form $\o$. Thus, we have a non-trivial observable,
\eqn\sympl{
W_{1,1}{}^{1,1}=\Fr{1}{4\pi^2}\int_M \tr(\w F^{1,1}
                 +\p\wedge\bar\p)\wedge\o\;.
}
One can see that $\hW_{1,1}{}^{1,1}$ is the K\"{a}hler form
${\tilde\o}$ on
$\CM^*$ descended symplectically from \kad, since
$F^{1,1}\wedge \o$ vanishes
for an instanton and the $\p$ and $\bar\p$ zero-modes represent the
holomorphic and the anti-holomorphic tangent vectors on $\CM^*$.
And, $\hW_{1,1}{}^{1,1}$ defines the Donaldson's $\m$-map,
\eqn\bao{
\eqalign{
\m(\o)&: H^{1,1}(M)\rightarrow H^{1,1}(\CM^*)\;,\cr
\m(\hbox{PD}[\o])&: H_{1,1}(M)\rightarrow H^{1,1}(\CM^*)\;,\cr
}}
where $\hbox{PD}[\o]$ denotes the homology class
which is Poincar\'{e} dual to $\o$.
Thus,  we have
\eqn\babo{
\left<\exp\left(\Fr{1}{4\pi^2}\int_M \tr\bigl(\w F^{1,1} +
\p\wedge\bar\p \bigr)\wedge \o\right)\right>
=\int_{\CM^*} \Fr{{\tilde{\o}}^d}{d!} = \hbox{vol}(\CM^*) > 0\;,
}
provided that $\CM^*$ is a non-empty compact manifold.
This argument is a field theoretical interpretation
of the Donaldson's differential geometric approach (
Theorem 4.1 as explained in p.~294--295 of
[2]). In practice, however,
the moduli space is rarely compact and it is hard to avoid
the singularities. These facts are both physically and mathematically
the main obstacle to understanding Donaldson theory more
fully [23]. For simply connected algebraic surfaces
under some suitable conditions,
Donaldson proved the positivity of the above invariant (Theorem 4.1 in
[2]) using the algebro-geometrical approach. We will sketch
his method in Sect.5.2.

\newsec{Other Approaches}
We have studied a natural field theoretical
interpretation of the Donaldson invariants on compact K\"{a}hler
surface.  However, it is not clear whether our
field theoretical method in general may give any new insight in
calculating the invariants explicitly.
We have seen that our formalism reduces
to the differential-geometrical approach of Donaldson.
On the other hand, Donaldson obtained
his positivity theorem by an independent
algebro-geometrical approach which is more
suitable in the realistic problems [2].
His method is
closely related to certain properties of the moduli space $\CM_\S^s$
of stable bundles over a compact Riemann surface $\S$.
In the language of stable bundle,
it is very natural to consider an $N=2$ TYMT on $\S$
which leads to a field theoretical interpretation of the
intersection pairings on $\CM_\S^s$. Moreover, Witten obtained
explicit expressions\fonote{Some similar results
was obtained by various mathematicians with different
methods [31].}, which is general enough to include arbitrary
compact group and the reducible connections, of the general intersection
pairings on the moduli space of flat connections on a Riemann
surface [4, 5]. It should be emphasized that his solutions
are based on field theoretical methods. On the other hand, the original
field theoretical interpretation of the Donaldson's invariants
has not produced any concrete result.
The main purpose of this final section is to
learn something useful to enhance the computability of the
Donaldson invariants from the Witten's solutions in two dimensions.
Although the TYMT in two dimensions was completely solved by Witten,
I will construct an $N=2$ version of the theory
to suggest a method, analogous to the Witten's solutions
in two dimensions, which may be useful in the calculation of the
Donaldson invariants on compact K\"{a}hler surfaces.

\subsec{$N=2$ TYMT on Compact Riemann Surfaces}
It is straightforward to obtain a two dimensional version of the $N=2$
TYMT or an $N=2$ version of the Witten's TYMT on Riemann surfaces (Sect.3
of [5]).
Let $\S$ be a compact oriented Riemann surface.  Picking a
complex structure $J$ on $\S$ with an arbitrary metric,
which is always a K\"{a}hler metric, we can determines a K\"{a}hler
form $\o$.  By the dimensional reasoning, every curvature two-form is
type $(1,1)$ and parallel to $\o$,
\eqn\flat{
F = f\o\;.
}
Thus, the constraint \lld\ is satisfied if $F$ is flat.  In this
sense a flat connection on Riemann surface can be regarded as
the two dimensional analogue of the anti-self duality relative to
K\"{a}hler form in compact K\"{a}hler surfaces.
Similarly, the self-dual two-forms $B,\c,\bar\c,H$
reduce to
\eqn\eaa{
B = B^{0}\o\;,\qquad
\c = \c^{0}\o\;,\qquad
\bar\c = \bar\c^{0}\o\;,\qquad
H = H^{0}\o\;.
}
The action of the two dimensional version of the $N=2$ TYMT
\eqn\eab{
S = \bs\bbs{\bf B_T}\;,
}
can be obtained from the same form of ${\bf B_T}$ given by Eq.\allegro
\eqn\ead{
{\bf B_T} = -\Fr{1}{h^2}\int_\S \tr\biggl[
\biggl(B^{0} f + \c^{0}\bar\c^{0}\biggr)
\biggr]\o\;.
}
The action is
\eqn\eae{\eqalign{
{\bf S}=&-\Fr{1}{h^2}\int_\S \tr\biggl[
H^{0}(H^{0}+f) +[\w,\c^{0}]\bar\c^{0}
+\bar\c^{0}\L \Dpp\p
+\c^{0}\L \Dp\bar\p \cr
&- B^{0}\L\left(\Dp \Dpp\w +[\p,\bar\p] +[\w,f]\right)
-\fr{1}{4}[\w,B^0][\w,B^0]
)
\biggr]\o\;.\cr
}
}
We can integrate $H^{0}$ out from the action
by setting $H^{0} =-\ha f$ or by the Gaussian integral,
which leads to modified transformation
\eqn\eaf{
\bs\bar\c^{0} = -\ha f+\ha[\w,B^0]\;,\qquad
\bbs\c^{0} =-\ha f-\ha[\w,B^0]\;.
}
Thus, the fixed point locus of $\bs$ and $\bbs$
is the moduli space flat connections with
the space of $\w$ zero-modes, modulo gauge symmetry.
The $\c^{0}$ and $\bar\c^{0}$
equations of motion,
\eqn\eag{
\qquad \L \Dpp\p =0\;,\qquad \L \Dp\bar\p =0\;,
}
together with the K\"{a}hler identities, show that the zero-modes
of $\p$ and $\bar\p$ are holomorphic and anti-holomorphic
tangent vectors on the moduli space $\CM_f$ of flat connections
on $\S$. If there is no reducible connection, the moduli space
is a smooth K\"{a}hler manifold with complex dimension being
equal to the number of the $\bar\p$ zero-modes. And the path
integral reduces an integral over the moduli space $\CM^*_f$ of
irreducible flat connections

It is also straightforward to construct observables.
Starting from $\CW_{0,0}{}^{2,2} = \ha\tr\w^2$, we can obtain
the essentially same relations as in Sect.4. The differences
are that $\CW_{p,q}{}^{r,s}$ is restricted to $p,q =0,1$,
$r,s=0,1,2$, and $F$ should be replaced by zero in the reduction
of $W_{p,q}{}^{r,s}$ to $\hW_{p,q}{}^{r,s}$.  Note that the
Hodge numbers on a compact oriented Riemann surface with genus
$g$ are given by
\eqn\eah{
h^{0,0}=h^{1,1}=1\;,\qquad
h^{1,0}=h^{0,1}=g\;.
}
In particular, the second cohomology group of $\S$
is represented by the K\"{a}hler form $\o$.
Then, the complete set of relevant observables is
\eqn\eai{\eqalign{
W_{0,0}{}^{2,2} &= \Fr{1}{8\pi^2}\int_\S \tr(\w^2)\o\;,\cr
W_{1,0}^{(g)2,1}&= \Fr{1}{4\pi^2}\int_{\S} \tr(\p\w)\wedge O_{0,1}^{(g)}\;,\cr
W_{0,1}^{(g)1,2}&= \Fr{1}{4\pi^2}\int_{\S} \tr(\bar\p\w)\wedge
O_{1,0}^{(g)}\;,\cr
W_{1,1}{}^{1,1}    &= \Fr{1}{4\pi^2}\int_\S \tr(\w F +\p\wedge\bar\p )\;.\cr
}
}
The observable
\eqn\eij{
\Fr{1}{4\pi^2}\int_\S \tr\bigl((\w F
+\p\wedge\bar\p\bigr)
}
reduces to  the K\"{a}hler form ${\tilde\o}_\S$
on moduli space of flat connections [27],
after replacing $F^{1,1}$ by zero and $\p,\bar\p$ by
their zero-modes. And, it gives the two-dimensional version of the
Donaldson's map
\eqn\aao{
\eqalign{
\m(1)&: H^{0,0}(\S)\rightarrow H^{1,1}(\CM^*_f)\;,\cr
\m(\hbox{PD}[1])&: H_{1,1}(\S)\rightarrow H^{1,1}(\CM^*_f)\;,\cr
}
}
Thus, the Donaldson's map actually induces the K\"{a}hler
structure $\tilde{\o}_\S$ on the moduli space $\CM^*_f$.  Consequently, the
correlation function,
\eqn\eik{
\left<\exp\left(\Fr{1}{4\pi^2}
\int_\S \tr\bigl(\w F +\p\wedge\bar\p \bigl)
\right)\right>
=\int_{\CM^*_f} \Fr{{\tilde{\o}_\S}^d}{d!} =\hbox{\rm vol}
(\CM^*_f)\;,
}
reduces to the symplectic volume of $\CM^*_f$, provided that
the moduli space is compact. This is analogous to Eq.\babo.

The intersection paring corresponding to \eik\
was  explicitly calculated by Witten [4]
by adopting various methods: i) zero coupling limit of physical
Yang-Mills theory; ii)  Verinde formula originated from the conformal
field theory; iii) relating combinatorial treatment of physical
Yang-Mills theory to the theory of Reidemeister-Ray-Singer torsion.
Note that the first and third methods of Witten are
related to physical Yang-Mills theory in two dimensions,
which also leads to a complete formalism\fonote{Witten obtained
a truly general expression which is valid regardless of
what properties
the moduli has, for an arbitrary Riemann surface.}
for the general intersection parings [5].
On the other hand, Verinde's formula is an algebro-geometrical formalism
although it has a physical origin [32].

\subsec{Algebro-Geometrical Approach}
We have seen that the correlation functions \babo\ and $\eik$
can be identified with the volumes of the moduli spaces
$\CM^*$ and $\CM^*_f$ respectively, provided that the moduli spaces
are compact. In this subsection, we will
sketch an independent algebro-geometrical method of Donaldson
on the intersection pairings [2], which is more suitable
to deal with the non-compactness of the moduli spaces.

Due to Donaldson,
a canonical identification of the correlation function can be
obtained by a theorem of Gieseker [33]
(and Proposition (5.4) in [2]).
Let $E \rightarrow \S$ be a complex vector bundle with reduction
of structure group $SU(2)$ over $\S$. From a fundamental
result of Quillen [34],
we have a determinant line bundle $\widetilde\CL_\S \rightarrow \CA_\S$
over the space of all connections on $E$. We  also have a $\CG$
invariant connection (Quillen metric) on $\widetilde\CL_\S$ with curvature
two form given by $-2\pi i$ times the K\"{a}hler structure
$\tilde{\o}_\S$ (Eq.(3.6) for $n=1$)
on $\CA_\S$. Then, we can obtain a holomorphic
line bundle $\CL_\S$ over the moduli space $\CM_f^*$ (or over
the moduli space $\CM_\S^s$ of stable bundles on $\S$)
with curvature given by $-2\pi i{\tilde\o}_\S$.
Thus, we have an another
representative of the Donaldson's map $\mu(PD[1])$
by the first Chern class, $c_1(\CL_\S)$,
of $\CL_\S$. We can also construct a line bundle $\CL_\S^{\otimes m}$.
The theorem of
Gieseker says that a section $j$
of some power $\CL_\S^{\otimes m}$ embeds  $\CM_\S^s$ as a
quasi-projective variety in $\BC\BP^N$.
The degree of a
quasi-projective variety $Y$ can be defined by the degree of
the closure $\overline Y$ in the projective space.
The degree deg$(\overline{j(\CM_\S^s)})$ of
the $d$-dimensional projective variety $\overline{j(\CM_\S^s)}$
is defined by the number of points of intersections of
$\overline{j(\CM_\S^s)}$ with a generic $(N-d)$ dimensional
hyperplane $\BC\BP^{N-d}\subset \BC\BP^N$. Equivalently
\eqn\degreee{
\hbox{\rm deg}(\overline{j(\CM_\S^s)})
=\left<c_1\left(\CL_\S^{\otimes m}\right)^d, \overline{j(\CM_\S^s)}
\right>\;.
}
Thus we can identify the intersection pairing represented by
the correlation function in \eik\ with $\hbox{\rm deg}
(\overline{j(\CM_\S^s)})/(d!\, m^d)$.

Furthermore,
Donaldson obtained his positivity theorem by constructing a
quasi-projective embedding $J$ of the moduli space of stable bundles on
a simply connected algebraic surface using
the Gieseker's projective embedding and
a theorem of Mehta and Ramanathan [35]. We will roughly describe
Donaldson's proof.
Let $M$ be a simply connected algebraic surface with $p_g(M)>0$
so that there is a Hodge metric
compatible with a projective embedding. Then, the cohomology class of
K\"{a}hler form $\o$ is Poincar\'{e} dual to the hyperplane section
class $H$. Let $\S$ be a Riemann surface representing $H$.
Now consider a moduli space $\CM^s_\S$ of stable bundles
over $\S$. Using the theorem of Mehta and Ramanathan, Donaldson
showed there is a restriction map $r: \CM_M^s \rightarrow \CM_\S^s$,
which is an embedding (for the precise statement, the reader should
refer to Sect.5 in [2]).
Over $\CM_\S^s$ we have a holomorphic line bundle $\CL_\S$ as mentioned
before. Then the Donaldson $\m(\S)$ map is the pull-back by $r$
of the first Chern class of $\CL_\S$. Now for large $m$ we have a
quasi-projective embedding $j(\CM_\S^s)$. Thus the composition $J=j\circ r$
gives a quasi-projective embedding of $\CM_M^s$. Then, for large $k$,
Donaldson showed that his invariant $q_{k,M}(H,\ldots,H)$ is proportional
to the $\hbox{deg}(\overline{J(\CM_M^s)})$.
Since the degree of a non-empty
variety is always positive, the positivity theorem is followed.

We will end this subsection by observing an
relation between th Donaldson $\m$-maps, \bao\ and \aao.
Let $M$ be a projective algebraic surface K\"{a}hler form $\o$ and $\S$
represent the hyperplane section.
Assume that we study the $N=2$ TYMT's
on $M$ and $\S$. The Donaldson $\m$-maps on $M$ and on $\S$
are represented by the topological observables,
\eqn\taag{\eqalign{
\Omega &= \Fr{1}{4\pi^2}\int_M \tr(\w F +\p\wedge\bar\p )\wedge\o\;,\cr
\Omega_\S &= \Fr{1}{4\pi^2}\int_\S \tr(\w F +\p\wedge\bar\p )\;.\cr
}}
The Poincar\'e duality
of $\o$ to $\S$ means that we have the following identity for every $(1,1)$
form $\eta$ on $M$ (p.~230 in [28]);
\eqn\taah{
\Fr{i}{2\pi}\int_M(\log h) \rd\bar\rd\eta = \int_\S\eta
-\int_M\eta\wedge\o\;,
}
where $h$ is a globally defined smooth function on $M$ which vanish exactly
at $\S$.
If we choose $\eta = \fr{1}{4\pi^2}\tr(\w F +\p\wedge\bar\p )$,
we have
\eqn\taai{
\O -\O_\S =
\bs\bbs\int_M(\log h)\,c_2(E)\;,
}
where we have used Eq.\taac\fonote{
Equation \taac\ also give a similar relation if choose $\eta =
\Fr{1}{8\pi^2}\tr(\w^2)\o$.
}. This is an interesting but not an accidental relation
as explained in Sect.6.5.4 in [7].
We want to construct a
holomorphic line bundle $\CL$ over $\CM_M^s$ with curvature
$-2\pi i$ times the K\"{a}hler form ${\tilde{\o}}=\widehat\O$ on
$\CM_M^s$, so that the first Chern class $c_1(\CL)$ of $\CL$ represents
the Donaldson $\m$-map \bao. The desired line bundle $\CL$ can be obtained
via pull back of the determinant line bundle $\CL_\S$
by the restriction map $r:\CM_M^s\rightarrow \CM_\S^s$.
Then, we get an induce connection
on $\CL$ with curvature form $-2\pi i\tilde{\o}_\S = -2\pi
i\widehat\O_\S$. Equation \taai\ implies that we can modify the pull
back connection by the zero-form $\int_M (\log h) c_2(E)$ on $\CA$ to get a
new connection which gives the desired curvature
$-2\pi i{\tilde{\o}}=\widehat\O$.
Since $\int_M (\log h) c_2(E)$ is gauge invariant, we have the
desired result.

\subsec{Non-Abelian Localization}
One of great challenges of the TYMT is to overcome its
formality and to provide some concrete computational methods.
In this view, Witten's beautiful results on two dimensions are
very encouraging [5]. The purpose of this subsection is
to suggest that the Witten's formalism can be applied to
calculating the Donaldson invariants on compact K\"{a}hler
surface.

To begin with, we briefly summarize Witten's
arguments\fonote{Note that the original arguments
of Witten are slightly different to what follows, since
the K\"{a}hler structure on Riemann surface is not essential
in his arguments. However, the K\"ahler structure is essential
to deal with the higher dimensional case.}.
Let $\S$ be a compact oriented Riemann surface with K\"{a}hler
form $\o$. The partition function of Yang-Mills
theory on $\S$ can be written as
\eqn\www{
Z(\e) = \Fr{1}{\hbox{vol}(\CG)}\int_{\CA_\S}\!\!\!\!
 \CD A\,\CD\p\,\CD\bar\p\,\CD\w\,
\exp\biggl[\Fr{1}{4\pi^2}\int_\S \tr\bigl(\w F +\p\wedge\bar\p\bigr)
+\Fr{\e}{8\pi^2}\int_\S \o\,\tr\w^2
\biggr]\;.
}
The decoupled fields $\p,\bar\p$ are introduced to give the correct
path integral measure (the symplectic measure in the K\"{a}hler
polarization). Witten's fundamental result is that
the partition function $Z(\e)$ of the two dimensional Yang-Mills
theory can be expressed as a sum of contributions of critical
points $S$
\eqn\kkb{
Z(\e) = \sum_{\a\in S} Z_\a(\e)\;.
}
In the limit $\e = 0$,
the only contribution of the critical points
is the absolute minimum of the action and the partition function
becomes an integral over the moduli space of flat connections.
For $\e\neq 0$, the higher critical points will contribute,
however, their contributions are exponentially small.  More
concretely,
\eqn\caa{
\biggl<\exp\biggl(\Fr{1}{4\pi^2}\int_\S \tr\left(\w F
+\p\wedge\bar\p\right)
+\Fr{\e}{8\pi^2}\int_\S \o\,\tr\w^2\biggr) \biggr>
= Z(\e) + O(\exp(-c)\;,
}
where $c$ is the smallest value of the Yang-Mills action,
$\Fr{1}{8\pi^2\e}\int_M \o\, \tr f^2$, for the higher
critical points. Using the correspondence \caa, Witten obtained
a general expressions for the intersection pairings on moduli
space of flat connections.

The above example should be viewed as an application of
the Witten's non-Abelian localization formula (a non-Abelian version
of the Duistermaat-Heckmann (DH) integral formula [36]) that a partition
function of a quantum field theory with an action functional given by
the norm squared of a moment map can be expressed as a sum of contributions
of critical points [5].
Now, we want to find a natural higher dimensional
analogue of the Witten's solution in two dimensions. We can start this
by observing some special properties of physical Yang-Mills
theory in two dimensions: i) every oriented compact Riemann
surface is a K\"{a}hler manifold; ii) every connection is
of type $(1,1)$ which defines a holomorphic structure; iii)
the action functional in the first order formalism, Eq.\www,
has the $N=2$ fermionic symmetry with the transformation law Eq.\symphony,
or the action functional consists of the topological observables
of the $N=2$ TYMT;
iv) the Yang-Mills action is given by the norm squared of the
moment map. Then, our strategy is to design a variant of physical
Yang-Mills theory on a compact K\"{a}hler surface $M$, which has
all the properties (i)-(iv).
An obvious candidate motivated from Eq.\www\ is
\eqn\wwwa{\eqalign{
Z(\e) = &\Fr{1}{\hbox{\rm vol}(\CG)}\int_{\CA^{1,1}}\!\!\!
\CD A\,\CD\p\,\CD\bar\p\,\CD\w\,
\exp\biggl[ \Fr{1}{4\pi^2}\int_M \tr\bigl(\w F
+\p\wedge\bar\p\bigr)\wedge \o  \cr
&+\Fr{\e}{8\pi^2}\int_M \Fr{\o^2}{2!}\tr\w^2
\biggr]\;,
}}
where we have restricted
$\CA$ to $\CA^{1,1}$. We further assume that $\p$ and $\bar\p$ are
tangent to $\CA^{1,1}$, i.e. $\Dp\p=\Dpp\bar\p=0$.
Similarly to its two-dimensional ancestor,
the decoupled fields $\p,\bar\p$ are introduced to give the symplectic
measure and to ensure the $N=2$ fermionic symmetry of the
theory. The above partition function is identical to
that of physical Yang-Mills theory on $M$ restricted to
$\CA^{1,1}$, which can be called {\it holomorphic Yang-Mills theory}
(HYMT), up to a topological term after integrating $\w,\p,\bar\p$ out.
And the action functional of the HYMT is the norm squared of the moment
map \taad. Then, by the non-Abelian version of thr DH integration formula
of Witten, it is obvious that the four-dimensional
counterpart of the mapping \caa\ exists between the intersection
pairings on the moduli space of ASD connections and the
partition function \wwwa. The only technical difficulty is to
define the restriction of $\CA$ to $\CA^{1,1}$ in a correct quantum
theoretical way. This can be easily achieved because the constraints
$F^{2,0}_A = F^{0,2}_A = \Dp\p=\Dpp\bar\p=0$, which define the HYMT,
are related by the $N=2$ fermionic symmetry. It will be more suitable
to define the HYMT as a deformation of the $N=2$ TYMT, analogous
to the simple mapping from the TYMT to physical Yang-Mills theory in
two dimensions [5]. The details will be discussed elsewhere [37].

\acknow{I am grateful to B.S.~Han, Dr.~S.~Hyun, G.H.~Lee and Dr.~ H.J.~ Lee
for discussions. Special thanks to Prof.~Q-Han Park for discussions and
reading on this manuscript.}

\bigbreak\bigskip\noindent{\it Note added}.
After submitting the first version of this paper,
the author receive a lecture notes
by G.~Thompson (Topological gauge theory and Yang-Mills theory,
Lecture notes in Trieste summer school, June 1992). In the final
section of the notes, he considered the symplectic volume of the
moduli space of ASD connections and a version of HYMT, both in terms of
$N=1$ TYMT on K\"{a}hler surfaces.

\bigbreak\bigskip
\centerline{\bf References}
\bigskip

\item{[1]\ }%
Witten, E.:
Topological quantum field theory.
\cmp {117}{1988}{353}
\item{[2]\ }%
Donaldson, S.K.:
Polynomial invariants for smooth $4$-manifolds.
\top {29}{1990}{257}
\item{[3]\ }%
Itoh, M.: On the moduli space of anti-self-dual Yang-Mills
connections on K\"{a}hler surfaces.
Publ.\ RIMS Kyoto Univ.\ {\bf 19}, 15 (1983);
The moduli space of Yang-Mills connections over a
K\"{a}hler surface is a complex manifold.
Osaka J.\ Math.\ {\bf 22}, 845 (1985);
Geometry of anti-self-dual connections and Kuranishi map.
J.\ Math.\ Soc.\ Japan, {\bf 40}, 9 (1988)
\item{[4]\ }%
Witten, E.:
On quantum gauge theories in two dimensions.
\cmp {141}{1991}{153}.
\item{[5]\ }%
Witten, E.:
Two dimensional gauge theories revisited.
\jgp {9}{1992}{303}.
\item{[6]\ }%
Galperin A., Ogievetsky, O.:
Holonomy groups, complex structures and $D=4$ topological Yang-Mills
theory.
\cmp {139}{1991}{377}
\item{[7]\ }%
Donaldson, S.K., Kronheimer, P.B.:
The geometry of four-manifolds. Oxford, New York:
Oxford University Press 1990
\item{[8]\ }%
Donaldson, S.K.: Yang-Mills Invariants of four-manifolds.
In: Geometry of low-dimensional manifolds $1$, Donaldson, S.K.,
Thomas, C.B. (eds.) (London Mathematical Society Lecture Note Series
$150$) London: Cambridge Unive.\ Press 1990
\item{[9]\ }%
Birminham, D., Blau, M., Rakowski, M., Thomson, G.:
Topological field theory. \prp {209}{1991}{129}
\item{[10]\ }%
Atiyah, M.F., Singer, I.M.:
Dirac operators coupled to vector potentials.
\pnas {81}{1984}{2597}
\item{[11]\ }%
Kanno H.: Weil algebraic structure and geometrical meaning of the BRST
transformation in topological quantum field theory.
\zp {C 43}{1989}{477}
\item{[12]\ }%
Atiyah, M.F., Hitchin, N.J., Singer, I.M.:
Self-duality in four dimensional Riemann geometry,
\ptrsls {A 362}{1978}{425}
\item{[13]\ }%
Freed, D.S., Uhlenbeck, K.K.: Instantons and four manifolds.
Berlin, Heidelberg, New York: Springer 1984
\item{[14]\ }%
Donaldson, S.K.:
The orientation of Yang-Mills moduli spaces and $4$-manifold topology.
\jdg {26}{1987}{397}
\item{[15]\ }%
Uhlenbeck, K.K.: Connections with $L^p$ bounds on curvature.
\cmp {83}{1982}{31}; Removable singularities in Yang-Mills fields.
\cmp {83}{1982}{11}
\item{[16]\ }%
Donaldson, S.K.:
Connections, cohomology and the intersection forms of four manifolds.
\jdg {24}{1986}{275}
\item{[17]\ }%
Baulieu, L., Singer, I.M.:
Topological Yang-Mills symmetry.
\np {(Proc.\ Suppl.) 5B}{1988}{12}
\item{[18]\ }%
Brooks, R., Montano, D., Sonnenschein, J.:
Gauge fixing and renormalization in topological quantum field
theory. \pl {B 214}{1988}{91}
\item{[19]\ }%
Labastida, J.M.F., Pernici, M.:
A gauge invariant action in topological quantum field
theory.
\pl {B 212}{1988}{56}
\item{[20]\ }%
Horne, J.:
Superspace versions of topological theories.
\np {B 318}{1989}{22}
\item{[21]\ }%
Myers, R.:
Gauge fixing of topological Yang-Mills.
\ijmp {A 5}{1990}{1369}
\item{[22]\ }%
Witten, E.:
The $N$ matrix model and gauged WZW models.
\np {B 371}{1992}{191};
Mirror manifolds and topological field theory.
IASSNS-HEP-91/83 \& hep-th/9112056.
\item{[23]\ }%
Witten, E.:
Introduction to cohomological field theories.
\ijmp {A 6}{1991}{2273}.
\item{[24]\ }%
Donaldson, S.K.:
Anti-self-dual Yang-Mills connections
on complex algebraic surfaces and stable bundles.
\plms {30}{1985}{1};
Infinite determinants, stable bundles and curvature.
\dmj {54}{1987}{231}
\item{[25]\ }%
Uhlenbeck, K.K., Yau, S.T.:
On the existence of hermitian Yang-Mills connections on stable bundles
over compact K\"{a}hler manifolds.
Commun.\ Pure \& Appl.\ Math.\
{\bf 39}, S257 (1986); Correction:
Commun.\ Pure \& Appl.\ Math.\
{\bf 42}, 703 (1987)
\item{[26]\ }%
Narasimhan, M.S., Seshadri, C.S.:
Stable bundle and unitary vector bundles on
compact Riemann surfaces.
\am {65}{1965}{540}
\item{[27]\ }%
Atiyah, M.F., Bott, R.:
The Yang-Mills equations over Riemann surfaces.
\ptrsls {A 308}{1982}{523}
\item{[28]\ }%
Kobayashi, S.: Differential geometry of complex vector bundle.
Publications of the mathematical society of Japan 15.
Princeton: Princeton University Press, 1987
\item{[29]\ }%
Wells, Jr., R.O.:
Differential analysis on complex manifolds.
Berlin, New York: Springer-Verlag Inc.\ 1980
\item{[30]\ }%
Kim, H.J.:
Moduli of Hermitian-Einstein vector bundles.
Math.\ Z.\ {\bf 195}, 143 (1987)
\item{[31]\ }%
Thaddeus, M.: Conformal field theory and the cohomology of the moduli
space of stable bundles. \jdg {35}{1992}{131}\semi
Kirwan, F.: The cohomology rings of moduli spaces of vector bundles over
Riemann surfaces. J.\ Amer.\ Math.\ Soc.\ {\bf 5}, 853 (1992)\semi
Donaldson, S.K.: Gluing techniques in the cohomology of moduli spaces.
To appear in Proceedings of the Andreas Floer memorial volume.
Hofer, H., Zehnder, E. (eds.)
\item{[32]\ }%
Verinde, E.: Fusion rules and modular transformations if $2$d conformal
field theory. \np {B 300}{1988}{360}
\item{[33]\ }%
Gieseker, D.: Geometric invariant theory and application to moduli
problem. In: Lecture notes in mathematics 996 (ed. Gherardelli, F.).
Berlin: Springer 1988
\item{[34]\ }%
Quillen, D.G.: Determinants of Caucy-Riemann operators over a Riemann
surface. Funct.\ Anal.\ Appl. {\bf 14}, 31 (1985)
\item{[35]\ }%
Metha, V.B., Ramanathan, A.: Restriction of stable sheaves and
representations of the fundamental group. \im {77}{1984}{163}
\item{[36]\ }%
Duistermaat, J.J., Heckmann, G.J.: On the variation in the cohomology
in the symplectic form of the reduced phase space. Invent Math.\ {\bf
69}, 259 (1982)
\item{[37]\ }%
Park, J.-S.: Holomorphic Yang-Mills theory on compact
K\"{a}hler surface. ESENAT-93-02 \& YUMS-93-11. Submitted to
Nucl.~Phys.~B

\end